\documentclass{aa}
\usepackage{graphicx}
\graphicspath{ {plots/} }
\usepackage{natbib}
\usepackage{txfonts}

\bibpunct{(}{)}{;}{a}{}{,} 

\def\kms{km~s$^{-1}$}
\def\teff{\textit{T}_{\text{eff}}}
\def\teffsun{\textit{T}_{\text{eff},\bigodot}}
\def\logg{\text{log}(\textit{g})}
\def\mh{[\text{M}/\text{H}]}
\def\feh{[\text{Fe}/\text{H}]}

\def\alpham{[\alpha/\text{M}]}

\def\snr{S/N}

\def\cms{cm~s$^{-2}$}

\begin{document}

\title{Beyond \emph{Gaia} DR3:\\Tracing the $\alpham-\mh$  bimodality\\ from the inner to the outer Milky Way disc\\ with \emph{Gaia}-RVS and convolutional neural networks\thanks{Full RVS-CNN catalog described in Table 2 is available via the AIP Gaia archive at https://gaia.aip.de/metadata/gaiadr3\_contrib/cnn\_gaia\_rvs\_catalog/. The query is done via the query interface https://gaia.aip.de/query/.}}

\author{G.~Guiglion \inst{1, 2, 3}, S.~Nepal \inst{3, 4},
C.~Chiappini \inst{3},
S.~Khoperskov \inst{3},
G.~Traven \inst{5},
A. B. A.~Queiroz  \inst{3},
M.~Steinmetz \inst{3},
M.~Valentini \inst{3},
Y.~Fournier \inst{3},
A.~Vallenari \inst{6},
K.~Youakim \inst{7},
M.~Bergemann \inst{2},
S.~M{\'e}sz{\'a}ros \inst{8,9}, 
S.~Lucatello \inst{10,11}, 
R.~Sordo \inst{6},
S.~Fabbro \inst{12},
I.~Minchev \inst{3},
G.~Tautvai{\v s}ien{\. e} \inst{13},
\v{S}.~Mikolaitis \inst{13}, 
J.~Montalb\'{a}n \inst{14}}

\institute{Zentrum f\"ur Astronomie der Universit\"at Heidelberg, Landessternwarte, K\"onigstuhl 12, 69117 Heidelberg, Germany
\and
Max Planck Institute for Astronomy, K\"onigstuhl 17, 69117, Heidelberg, Germany.
\and
Leibniz-Institut f\"ur Astrophysik Potsdam (AIP), An der Sternwarte 
16, 14482 Potsdam, Germany \\
\email{guiglion@mpia.de}
\and
Institut f\"ur Physik und Astronomie, Universit\"at Potsdam, Karl-Liebknecht-Str. 24/25, 14476 Potsdam, Germany \\
\email{snepal@aip.de}
\and 
Faculty of Mathematics and Physics, University of Ljubljana, Jadranska 19, 1000 Ljubljana, Slovenia
\and
INAF, Osservatorio di Padova, Vicolo Osservatorio 5, 35122, Padova
\and
Department of Astronomy, Stockholm University, AlbaNova University Centre, Roslagstullsbacken, 106 91 Stockholm, Sweden
\and
ELTE E\"otv\"os Lor\'and University, Gothard Astrophysical Observatory, 9700 Szombathely, Szent Imre H. st. 112, Hungary
\and
MTA-ELTE Lend{\"u}let "Momentum" Milky Way Research Group, Hungary
\and
INAF–Osservatorio Astronomico di Padova, Vicolo dell’Osservatorio 5, I-35122 Padova, Italy
\and
Institute for Advanced Studies, Technische Universit\"at M\"unchen, Lichtenbergstraße 2a, D-85748 Garching bei M\"unchen, Germany
\and
National Research Council Herzberg Astronomy \& Astrophysics, 4071 West Saanich Road, Victoria, BC, Canada
\and
Institute of Theoretical Physics and Astronomy, Vilnius University, Sauletekio av. 3, 10257, Vilnius, Lithuania
\and
Dipartimento di Fisica e Astronomia, Universit\`a di Bologna, Via Gobetti 93/2, I-40129 Bologna, Italy}

\date{Received 08/06/2023 ; accepted 23/10/2023}

\abstract
{In June 2022, \emph{Gaia} DR3 has provided the astronomy community with about one million spectra from the Radial Velocity Spectrometer
(RVS) covering the CaII triplet region.
In the next \emph{Gaia} data releases, we anticipate the number of RVS spectra
to successively increase from several 10 million spectra to eventually
more than 200 million spectra. Thus, stellar spectra are projected to be produced on an
`industrial scale', with numbers well above those for current and anticipated
ground-based surveys. However, one-third of the published spectra have $15\le\snr\le25$ per pixel
such that they pose problems for classical spectral
analysis pipelines, and therefore,
alternative ways to tap into these large datasets need to be devised.}
{We aim to leverage the versatility and capabilities of machine learning
techniques for supercharged
stellar parametrisation by combining \emph{Gaia}-RVS spectra with the full set of
\emph{Gaia} products and high-resolution, high-quality ground-based spectroscopic
reference datasets.}
{We developed a hybrid convolutional neural network (CNN) that combines the \emph{Gaia}
DR3 RVS spectra, photometry (G, $\mathrm{G}\_{\mathrm{BP}}$, $\mathrm{G}\_{\mathrm{RP}}$), parallaxes, and XP coefficients to derive atmospheric
parameters ($\teff$, $\logg$ as well as overall $\mh$) and chemical abundances
($\feh$ and $\alpham$). We trained the CNN
with a high-quality training sample based on APOGEE DR17 labels.}
{With this CNN, we derived homogeneous atmospheric parameters and abundances for 886\,080 RVS stars that show remarkable precision and accuracy
compared to external datasets (such as GALAH and asteroseismology).
The CNN is robust against noise in the RVS data, and we derive very precise labels down to S/N=15.  We managed to
characterise the $\alpham-\mh$ bimodality from the inner regions to the outer
parts of the Milky Way, which has never been done using RVS spectra or
similar datasets.}
{This work is the first to combine machine learning with such diverse datasets
and paves the way for large-scale machine learning analysis of \emph{Gaia}-RVS
spectra from future data releases.
Large, high-quality datasets can be optimally combined thanks to the CNN,
thereby realising the full power of spectroscopy, astrometry, and photometry.}

\keywords{Galaxy: stellar content - stars: 
abundances - techniques: spectroscopic - method: data analysis}

\titlerunning{Raving with \emph{Gaia}-RVS}
\authorrunning{G. Guiglion, S. Nepal et al.}

\maketitle



\section{Introduction}

Precise stellar chemical abundances are crucial to constraining the formation and evolution 
of the Milky Way and its neighbouring galaxies, as they allow stars to be used as
fossil records of past star formation events and enable the disentangling of stellar populations
or the tracing of accreted stars and stellar streams (e.g. \citealt{Matteucci2021, Helmi2020}). The stellar elemental abundances, coupled with
astrometry from the \emph{Gaia} space mission (e.g. \citealt{GaiaCollaboration2016, lindegren2018,
Lindegren2021}), are the chemo-dynamical process that
shaped the Milky Way and its satellites we observe today (e.g. \citealt{Tolstoy2009, Bergemann2018,
Haywood2018, Queiroz2021}). The more detailed the stellar chemistry is, the more we can
know about the nucleosynthesis processes that occurred (e.g. \citealt{Nomoto1997,
Roederer2016, anders_2018}). From an observational point of view, this translates
into the necessity of measuring a large variety of spectral lines from many different
elements in stellar spectra.

Deriving high-precision chemical abundances for Galactic Archaeology
has become a quest for modern large-scale spectroscopic surveys. Surveys
use modern spectrographs to observe stars with different setups and at
different spectral resolutions. For instance, the \emph{Gaia}-ESO survey (GES;
\citealt{Gilmore2022, Randich2022}) used both the ESO UVES and GIRAFFE
high-resolution spectrographs covering large wavelength ranges (from near-UV to
near-IR) and observing at different resolutions (from 16\,000 up to 48\,000) 
about $10^5$ stars.
Other surveys, such as LAMOST, have followed the same approach at
low and intermediate resolution \citep{zhang2019, Zhang2020, Wang2020} and
targeted almost $8\times10^8$ stars.
Additionally, infrared spectroscopy is very important for gathering spectra from
stars located in high-extinction regions. The Apache Point Observatory Galactic Evolution Experiment (APOGEE) pursued this effort ($R=22\,500$,
$\lambda \in 1.5-1.7\mu$m; \citealt{apogeedr16, Abdurrouf2022}) 
by observing more than $600\,000$ stars, contributing greatly to furthering our
understanding of the formation and evolution of the Galactic bulge (e.g. \citealt{Rojas2019,
queiroz2020}). The goal  of the WEAVE \citep{WEAVE} and 4MOST \citep{4MOST} surveys will be to respectively observe the northern and southern hemispheres at both
low-($R\sim5\,000$) and high-($R\sim20\,000$) resolution over the optical domain.
Those surveys will have to deal with an unprecedented number of spectra ($>10^7$) in their ultimate
goal of measuring numerous (>15 elements) high-quality abundances
\citep{bensby2019, chiappini2019, Christlieb2019, Helmi2019, Cioni2019, Jin2023}.

Originally, standard spectroscopy was the best way to derive a large variety of
chemical abundances in stellar atmospheres. Large spectroscopic surveys base their spectral analysis on standard spectroscopic techniques, which are based on the knowledge
of stellar atmospheric properties (such as pressure, temperature, and density;
see \citealt{Gray_2005oasp.book}) and radiative transfer (emission and
absorption mechanisms in the stellar atmosphere). A detailed knowledge of spectral absorption
lines is also needed (e.g. \citealt{Guiglion2018RNAAS, heiter2021, Kordopatis2023})
in order to find the best features for chemical abundance derivation. Departure
from local thermodynamic equilibrium can influence abundance determination (e.g.
\citealt{Bergemann2012}). Spectral fitting techniques, equivalent width methods,
and differential spectroscopy are usually employed to extract abundances and atmospheric
parameters from an observation, for instance, SME \cite{valenti1996}, GAUGUIN
\citep{guiglion2016}; however, Bayesian methods have also been used \citep{schonrich2014, Gent2022}. Notably, this type
of method has been intensively used over the last two decades, and it still continues to
play a crucial role in the analysis of modern survey data.

\begin{figure}[]
        \centering
        \includegraphics[width=0.99\linewidth]{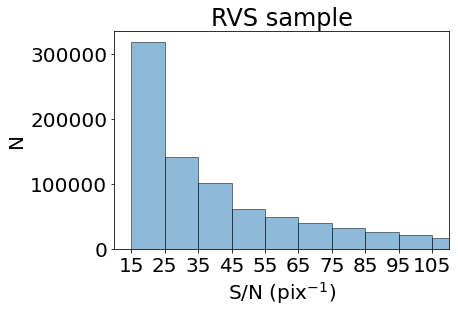}
        \caption{Signal-to-noise ratio distribution of the RVS sample used in this study.}
\label{fig:snr_distribution}
\end{figure}

\begin{figure}[]
        \centering
        \includegraphics[width=\linewidth]{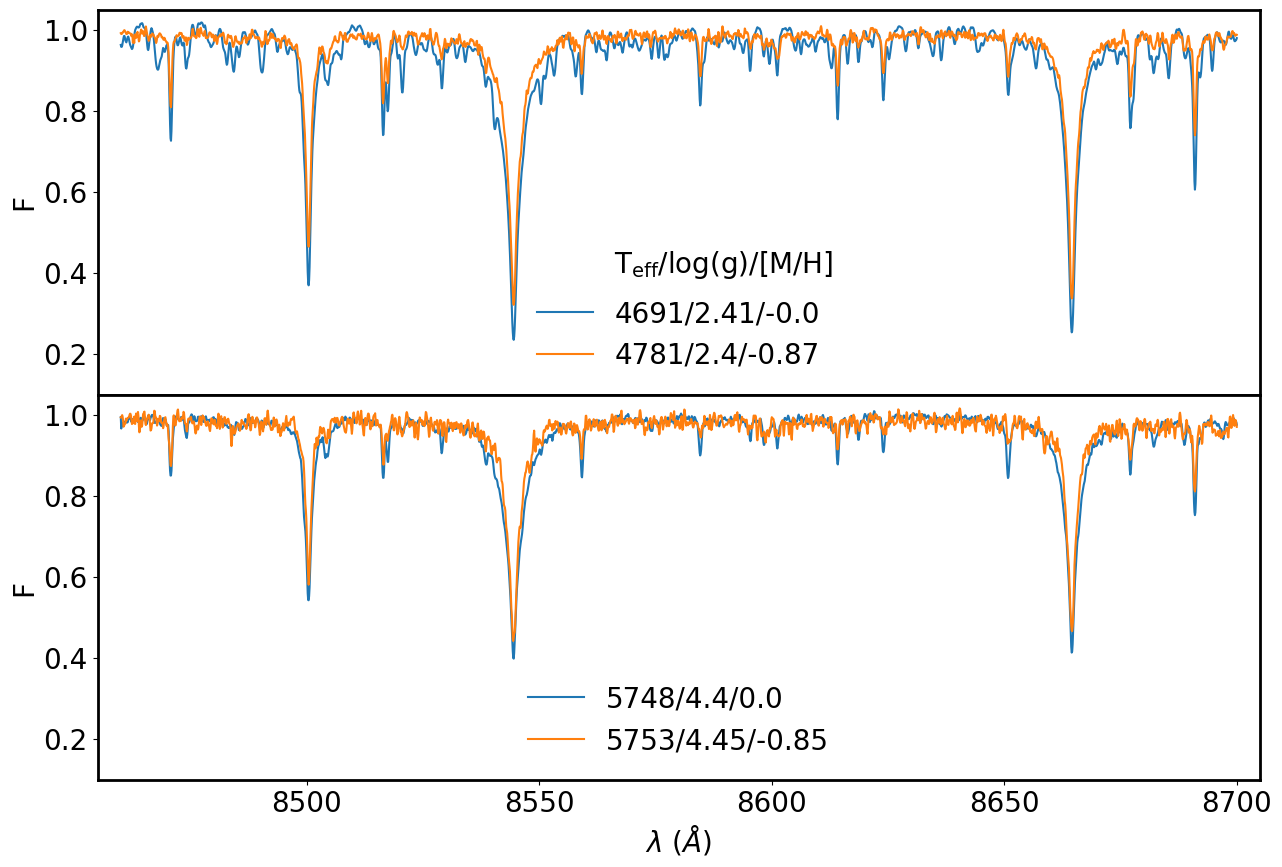}
        \caption[]{Examples of \emph{Gaia}-RVS spectra for an RC star (top) at $\mh=0$
 (blue) and $\mh=-0.87\,$dex (orange). The bottom panel shows a solar twin with
 two different metallicities.}
        \label{fig:rvs_spec}
\end{figure}

\begin{figure}[]
        \centering
        \includegraphics[width=0.99\linewidth]{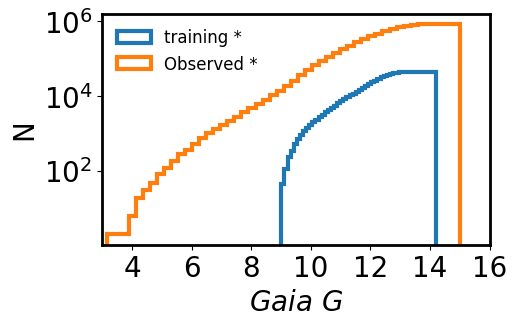}
        \caption{Density distribution of \emph{Gaia} G magnitudes
 in the training (44\,780 stars, blue) and observed (841\,300, orange)
 samples. Approximately $98\%$ of the observed magnitudes are contained within
the training sample limits.}
\label{fig:Gmag_histo}
\end{figure}

\begin{figure}[]
        \centering
        \includegraphics[width=\linewidth]{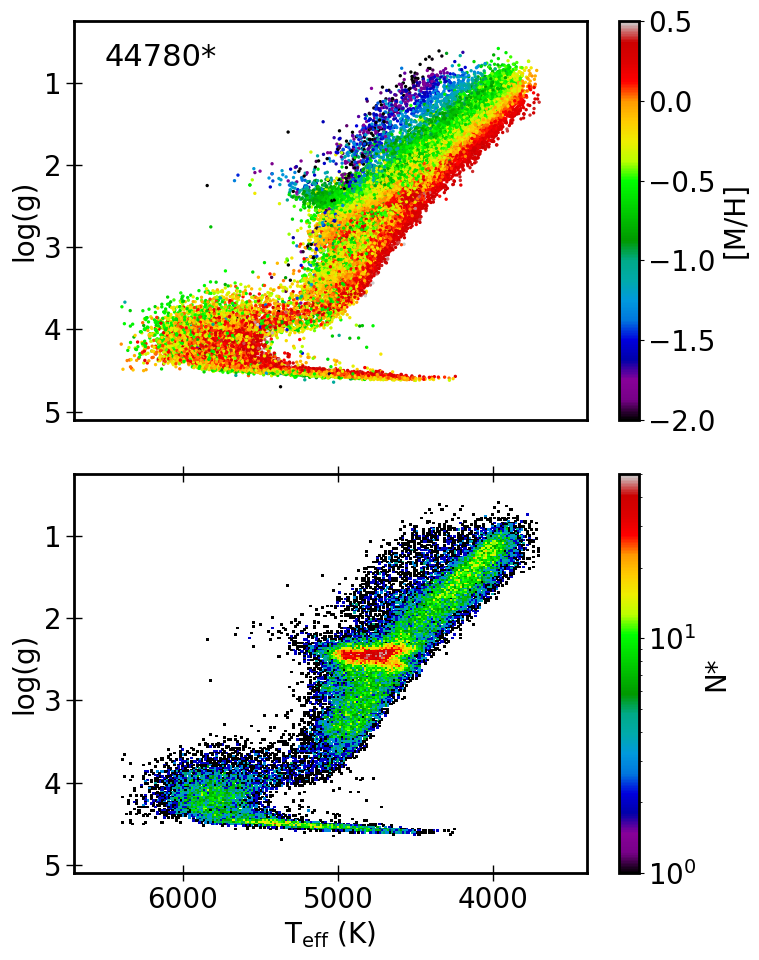}
        \caption{Kiel diagram of the APOGEE DR17 input labels of the
 training sample (44\,780 stars) coloured by $\mh$
 (top) and in density plot fashion (bottom). Our training sample is clearly
 dominated by giants (77\% of the sample has $\logg$<3.5), which is a direct
 effect of both the APOGEE and \emph{Gaia}-RVS selection functions.}
        \label{fig:training_sample_kiel}
\end{figure}

New, extremely large spectral surveys have pushed standard spectroscopic techniques to their limit, requiring a fundamental shift in
the spectral analysis methods.
Machine learning (ML) methods have been used to propagate the knowledge of standard spectroscopy to large-scale
spectroscopic datasets. The main idea is to build a set of reference stars (training
sample) with atmospheric parameters and chemical abundances (stellar labels) determined
using standard spectroscopy. An ML model is then built between stellar spectra and stellar
labels and propagated to an external set of spectra. Such a method is powerful
because it allows for the simultaneous derivation of many stellar labels for millions of spectra,
typically in several minutes. Meticulous selection of the training sample is the crucial
part of an ML framework, as biased training samples automatically lead to biases in
the label prediction. Several types of ML algorithms have recently been employed in the
Galactic Archaeology community in order to parameterise stellar spectra. Examples include  the Cannon
algorithm \citep{ness2015}, which builds a generative model between spectra and stellar labels; the
Payne algorithm \citep{ting2019}, which is based on the same generative model framework
but contains an interpolator based on an artificial neural network
(ANN) that uses synthetic spectra ab-initio; ANNs \citep{BailerJones1997}, and convolutional neural networks (CNNs), which
build a model between spectra and labels as well. Notably, CNNs are extremely efficient in learning from spectral features,
for example, AstroNN \citep{leung2019a} and StarNet \citep{fabbro2018, Bialek2020}. Concepts
and details on CNNs can be found in \citet{LeCun_backpropagation_1989} and
\citet{lecun-bengio-95a}.

In \citet{guiglion2020}, we showed that it was possible to extract precise chemical
information from spectra with limited resolution and wavelength coverage. We derived
homogeneous atmospheric parameters and chemical abundances from the spectra from the sixth
data release (DR6) of the RAdial Velocity Experiment (RAVE; \citealt{steinmetz2020a,
Steinmetz2020b}). We used a CNN approach trained on stellar labels from the 16th Data
Release (DR16) of APOGEE (\citealt{Ahumada2020}). This helped alleviate some of the spectral degeneracies inherent
to the RAVE spectra ($R\sim7500$, $\lambda\in\,842\,0-8\,780$\,\AA) by using
complementary absolute magnitudes computed from 2MASS, ALLWISE, and \emph{Gaia} DR2
photometry and parallaxes \citep{babusiaux2018, lindegren2018}. The uncertainties on
the resulting atmospheric parameters and chemical abundances were two to three times lower
than those reported by RAVE DR6. Including such extra constraints in the form of photometry
and parallaxes has also been done recently in the context of APOGEE for
parameterising together low- and high-mass stars with a CNN \citep{sprague2022}.

Recently, \citet{Nepal2023} and \citet{Ambrosch2023} achieved major improvements on the use of CNNs
for chemical abundances using GES spectra, including more complex and refined architectures, improved training strategies
and uncertainties, and improved reliability and robustness with respect to
radial velocities, rotational velocities, and signal-to-noise. Both studies showed
that CNNs efficiently learn from spectral features instead of abundance correlations,
which is key for detecting chemically peculiar stars (e.g. lithium-rich giants;
\citealt{Nepal2023}). These two studies represent a major step forward in the comprehension
of CNNs for the exploitation of future surveys, such as 4MOST and WEAVE.

\begin{figure*}[h]
        \centering
        \includegraphics[width=0.99\linewidth]{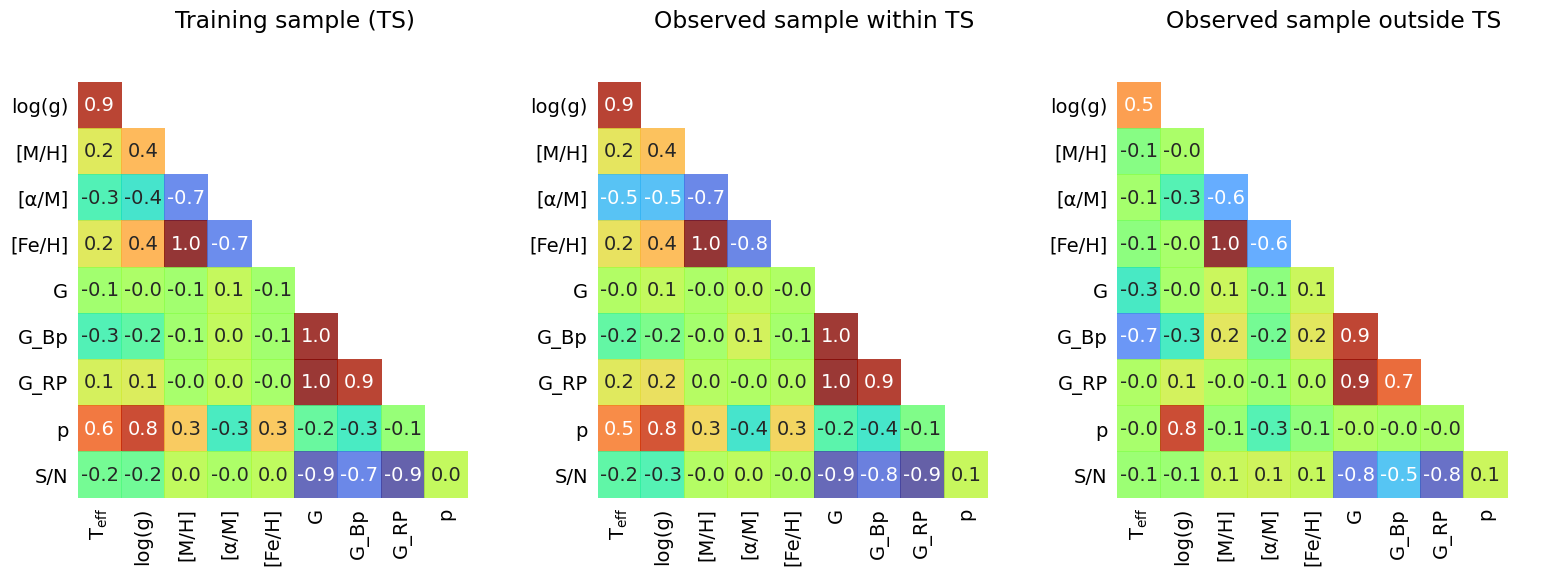}
        \caption[heatmap]{Correlation matrix of the labels together with
 photometry (\emph{Gaia} G, $\mathrm{G}\_{\mathrm{BP}}$, and $\mathrm{G}\_{\mathrm{RP}}$), parallaxes (p), and S/N.
 Left: Training sample (44\,780 stars). Middle: Observed sample within training sample limits (644\,287 stars). Right: Observed sample outside of the training sample limits (197\,013 stars). The only parameter not fed to the CNN during training was the S/N.}
        \label{fig:correlation_matrix_training_set}
\end{figure*}

In June 2022, the \emph{Gaia} consortium released around 1 million epoch-averaged RVS
spectra that were originally analysed during \emph{Gaia}
DR3 (10.17876/gaia/dr.3) by the General Stellar Parametriser
for spectroscopy (GSP-Spec; \citealt{recioblanco2023}) module of the Astrophysical parameters
inference system (Apsis; \citealt{Creevey2023}). Among these 1 million spectra,
one-third have $15\le\snr<25$ (see \figurename~\ref{fig:snr_distribution}),
for which GSP-Spec did not provide atmospheric parameters nor $\alpham$ ratios with `good' flags\_gspspec.\footnote{By `good', we refer to the first 13 flags of the flags\_gspspec chain equal to `zero' (see Table 2 \citealt{recioblanco2023}). Such flags make sure that, for instance, the parameters have an accuracy better than 250K in $\teff$, 0.5 in $\logg$, and 0.25 in $\mh$ or that no emission features or negative flux values could hamper the GSP-Spec parameterisation.} The main aim of this work is to
obtain precise atmospheric parameters ($\teff$, $\logg$, $\mh$) and chemical abundances
($\feh$, $\alpham$) down to S/N=15 for the \emph{Gaia} DR3 RVS spectra so that new science studies can leverage the larger, higher-quality
dataset. To achieve this
goal, we combined a hybrid CNN approach using APOGEE DR17 stellar labels with RVS spectra,
photometry (G, $\mathrm{G}\_{\mathrm{BP}}$, $\mathrm{G}\_{\mathrm{RP}}$), parallaxes, and
XP coefficients in order to break the
spectral degeneracies. For the first time, the precise chemistry derived with a CNN allowed us to trace the $\alpham-\mh$ bimodality with \emph{Gaia}-RVS data.

The paper is divided as follows: In Sect.~\ref{training_sample}, we present the dataset
used and the creation of the training sample. In Sect.~\ref{cnn_method}, we detail the CNN
method we used. In Sect.~\ref{prediction_observed_sample}, we present the parameterisation
of the \emph{Gaia}-RVS spectra. In Sect.~\ref{outside_ts_limits} we provide a way to ensure that a CNN label is within the training sample limits, while in Sect~\ref{validation_section} we validate our CNN labels with external datasets. In Sect.~\ref{science_verification}, we trace the
$\alpham-\mh$ bimodality in the Milky Way disc, and we list some caveats and draw
conclusions in Sect.~\ref{Caveats} and Sect.~\ref{conclusiooooonnnn}, respectively.


\section{Data}\label{training_sample}

The data used in the present study consists of \emph{Gaia} DR3 RVS spectra
\citep{gaiadr3_survey_properties}. We also incorporated \emph{Gaia}
DR3 photometry (phot\_g\_mean\_mag G; phot\_bp\_mean\_mag $\mathrm{G}\_{\mathrm{BP}}$;
and phot\_rp\_mean\_mag $\mathrm{G}\_{\mathrm{RP}}$ magnitudes; \citealt{Riello2021}), parallaxes
\citep{Lindegren2021}, and XP coefficients \citep{denageli_2023}.
The labels
of the training sample are from APOGEE DR17 \citep{Abdurrouf2022}.

\begin{figure*}[h]
        \centering
        \includegraphics[width=0.99\linewidth]{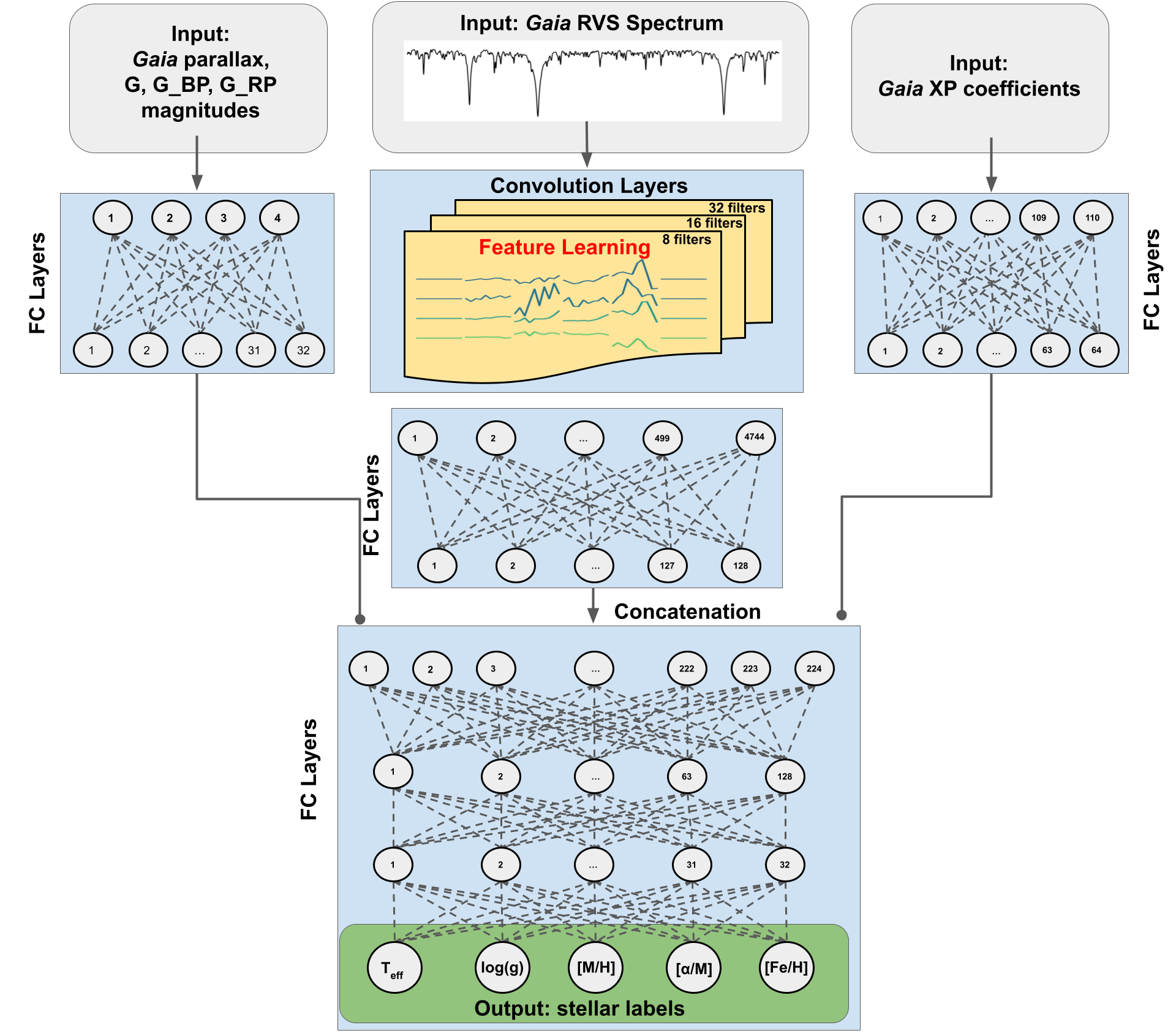}
        \caption[one\_to\_one]{Flow chart of CNN. \emph{Gaia}-RVS spectra
 are used as input spectra and passed through the convolution layers
 for feature extraction. Extra information is fed to the CNN in the form
 of \emph{Gaia} DR3 parallaxes as well as G, $\mathrm{G}\_{\mathrm{BP}}$, 
 and $\mathrm{G}\_{\mathrm{RP}}$ photometry (on the
 left) and XP coefficients (on the right). The information is
 fully connected (FC) to the output labels $\teff$, $\logg$, $\mh$,
 $\alpham$, and $\feh$.}
        \label{fig:cnn_architecture}
\end{figure*}

\subsection{\emph{Gaia}-RVS spectra}

We used the $999\,670$ time-averaged, normalised, and radial-velocity corrected
\emph{Gaia}-RVS spectra\footnote{https://doi.org/10.17876/gaia/dr.3/54}
from \emph{Gaia} DR3 (Seabroke et al., 2022).
The \emph{Gaia}-RVS spectra contain 2\,401 pixels along a scan,
with a pixel size of 0.10\AA \ and covering a spectral range of
8\,460~to~8\,700\,\AA ~(240\,\AA\, range). The spectral resolving power
reported by the \emph{Gaia} consortium is $R\sim11\,500$ \citep{Katz2023}.
Using the flags present in the \emph{Gaia} DR3 archive,
we removed potential galaxies and quasars (in\_galaxy\_candidates =
False and in\_qso\_candidates = False) as well as objects showing
variability (phot\_variable\_flag $\ne$ 'VARIABLE') and binarity signs
(non\_single\_star = 0). 
As some of the \emph{Gaia}-RVS may contain NaN values, we replaced such values by an
upper quartile (Q3) of continuum flux computed in three regions
(8\,475-8\,495\,\AA, 8\,561-8\,582\,\AA, and  8\,627-8\,648\,\AA).
These regions were selected to not contain any strong
spectral features, but they are not completely devoid of lines.
In \figurename~\ref{fig:snr_distribution},
we show a S/N distribution of the $Gaia$-RVS sample spectra. One can clearly see that the sample 
is dominated by spectra with $15\le\snr\le25$, as they comprise one-third of the sample.
Examples of red clump (RC) stars and solar twin
spectra are presented in Fig.~\ref{fig:rvs_spec}. In addition to
the strong \ion{Ca}{II} triplet, RVS spectra contain a large variety
of smaller features (see Sect.~\ref{section_gradients}).

\subsection{\emph{Gaia} DR3 magnitudes and parallaxes}

In addition to \emph{Gaia}-RVS spectra, we adopted \emph{Gaia}
DR3 magnitudes G (330-1050 nm), $\mathrm{G}\_{\mathrm{BP}}$ (330-680 nm), and
$\mathrm{G}\_{\mathrm{RP}}$ (630-1050 nm)
and parallaxes.\footnote{https://doi.org/10.17876/gaia/dr.3/1} When combined with magnitudes, parallaxes give information on the
luminosity and thus on the surface gravity and temperature of stars. We
removed spurious magnitudes by applying
phot\_[g/bp/rp]\_mean\_flux\_over\_error$>$500.
We removed negative parallaxes as well. Contrary to
\citet{guiglion2020}, we did not
compute absolute magnitudes in order to give the CNN more flexibility.
We also note that we did not apply parallax corrections
as described in \citet{Lindegren2021}.

\subsection{\emph{Gaia} DR3 XP coefficients}

\emph{Gaia} DR3 also provides low-resolution~($R\sim$30-100)
time-averaged spectra\footnote{https://doi.org/10.17876/gaia/dr.3/53}
in the blue (BP) and the red (RP) for 220 million
stars \citep{denageli_2023}). These so-called XP spectra have been
intensively used within the \emph{Gaia} collaboration for deriving among other
photometric effective temperatures, surface gravities, and metallicities
(GSP-Phot pipeline; \citealt{Andrae2023a}). The XP are also known to contain
metallicity-sensitive features \citep{Andrae2023b, Zhang2023, Yao2023, Xylakis2022}.
The XP spectra are given
in the form of a projection onto a set of
basis functions
(i.e. the coefficients of the projection; \citealt{denageli_2023}) to complement
RVS spectra, parallaxes, and $G/\mathrm{G}\_{\mathrm{BP}}/\mathrm{G}\_{\mathrm{RP}}$ photometry. \emph{Gaia} provides
55 BP and 55 RP coefficients (i.e. 110 XP coefficients). The XP coefficients give the CNN more
features to learn the atmospheric parameters and abundances we aim to derive.
We required that a given RVS spectrum has
available XP coefficients (has\_xp\_continuous=True).
To filter emission, we applied classlabel\_espels == NaN
\citep{Creevey2023}.

\begin{figure*}[h]
        \centering
        \includegraphics[width=\linewidth]{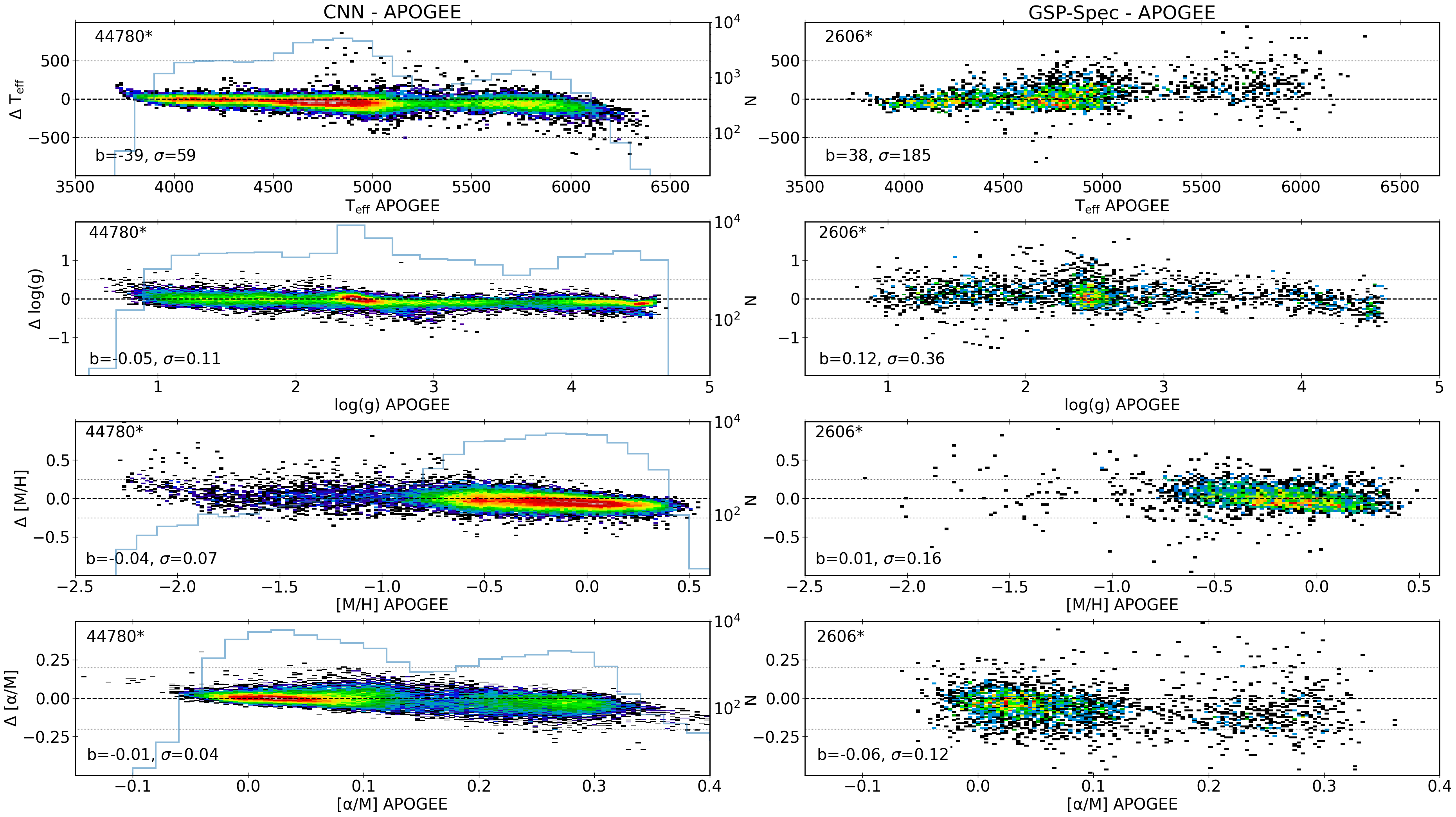}
        \caption[]{Two-dimensional density distribution
 of the residual between trained CNN labels and APOGEE input labels as
 a function of the APOGEE input labels for the training sample
 (44\,780 stars, left column). The black dashed line shows a null difference.
 The mean bias (b) and dispersion ($\sigma$) of the difference
 is given in the bottom-left
 corner. Each panel also contains a histogram of the input label. The right panels show differences between
 the calibrated GSP-Spec parameters
 (with good quality flags; \citealt{recioblanco2023}) and APOGEE labels.}
        \label{fig:training_vs_gspspec}
\end{figure*}

\subsection{Labels of the training sample}\label{labels_train_set}

In this paper, we aim at deriving the main atmospheric parameters
$\teff$, $\logg$, and overall $\mh$ together with the Fe content $\feh$
and $\alpham$. As stellar labels for the training sample, we used
the high-quality calibrated atmospheric parameters and
individual chemical
abundances from the 17th Data Release (DR17) of APOGEE
\citep{Abdurrouf2022}, which contains $733\,901$ stars.
This dataset is the best suited
for our CNN application, as APOGEE DR17 covers both the northern and
southern hemispheres (as does \emph{Gaia}), thus maximising the size of
the training sample. We cross-matched APOGEE DR17 and
\emph{Gaia}-RVS data based on \emph{Gaia} EDR3
source\_id, leading to a crossmatch of $207\,953$ stars.
We also filtered duplicates based on \emph{Gaia} source\_id.
We extensively used the flags
provided by APOGEE DR17 and followed the recommendations of the
survey to clean up the sample in order to make it the most reliable it can be.
From the APOGEE\_ASPCAPFLAG bitmask, we used the Binary Digits 7 (STAR\_WARN)
and 23 (STAR\_BAD), that is, removing stars showing potentially
bad $\teff$ and $\logg$, large CHI2, a large discrepancy between
the infrared flux method and spectroscopic temperatures,
systematics due to large rotation, and S/N<70 (see
\citet{Abdurrouf2022} and sdss.org/dr17/irspec/apogee-bitmasks/
for more details). We also selected stars
with ASPCAP\_CHI2<25. We selected stars with
$\feh$ flag E\_H\_FLAG=0. To remove possible spurious
measurements, we performed a cut in atmospheric parameters and abundances:
TEFF\_ERR$<$100\,K, LOGG\_ERR$<$0.1 dex, M\_H\_ERR$<$0.2 dex if
M\_H$<$-0.5 dex and M\_H\_ERR$<$0.1 dex if M\_H$>$-0.5 dex (same
condition for $\feh$), and ALPHA\_M\_ERR<0.1 dex. We note that
APOGEE DR17 $\alpham$ was derived thanks to a mixture of Ca, Ti, Mg, Si, O, Ne,
and S lines in APOGEE spectra\citep{Abdurrouf2022}, while RVS spectra has several
Ca, Ti, and Si lines relevant for $\alpham$ derivation (see Sect~\ref{rvs_gradients} for more
details)

\subsection{Final training and observed samples}

In order to build the training sample, we selected the RVS spectra with
the corresponding labels detailed in Section~\ref{labels_train_set}.
Tests have shown that a too low S/N would degrade the
learning performances of our CNN method \citep{guiglion2020, Nepal2023};
hence we adopted \emph{Gaia}
DR3 rvs\_spec\_sig\_to\_noise ratios larger than 30
(S/N$\ge$30 pix$^{-1}$) for the training sample spectra.
We emphasise that  rvs\_spec\_sig\_to\_noise is
defined as the signal-to-noise ratio of the mean RVS spectrum.
For the training sample only, we limited our sample to have parallax absolute
errors lower than 20\% and RUWE<1.4 (i.e. stars with a good single star astrometric solution).

Following the series of cuts, we had
in hand a training sample composed of $44\,780$ \emph{Gaia}-RVS spectra
together with their respective \emph{Gaia} G, $\mathrm{G}\_{\mathrm{BP}}$, and 
$\mathrm{G}\_{\mathrm{RP}}$ pass-bands,
parallaxes, XP coefficients, and APOGEE DR17 stellar labels $\teff$,
$\logg$, $\mh$, $\feh$, and $\alpham$. 

The rest of the \emph{Gaia}-RVS spectra (N = 841\,300, with no training sample
labels but with parallaxes; G, $\mathrm{G}\_{\mathrm{BP}}$, and $\mathrm{G}\_{\mathrm{RP}}$
photometry; and XP coefficients) constitute
the `observed sample' that we aimed to parameterise with the CNN.

\subsection{Dynamical range of the training sample}

\tablename~\ref{label_range_training_sample}
contains the physical range of the stellar labels. In
\figurename~\ref{fig:training_sample_kiel}, we show a Kiel diagram of the
training sample. The giants ($\logg<3.5$) represent 70\% of the training
sample. The metal-poor tail ($\mh<-1$) is composed of 1578 stars, which
leads to very reliable metallicities down to $-2.3$\,dex. We note that in
\citet{guiglion2020}, the training sample was ten times smaller and only
included 70 stars with $\mh<-1$. In \figurename~\ref{fig:Gmag_histo},
we show the distribution of \emph{Gaia} DR3 G magnitudes in the
training and observed samples. The training sample ranges from $7 < G < 14.2$,
while the observed set includes brighter (down to $G=3.2$) and fainter
(up to $G=15$) targets. We note that 98\% of the observed sample is included within the magnitude range covered by the training sample.

In \figurename~\ref{fig:correlation_matrix_training_set}, we show a
correlation matrix of labels, photometry, parallaxes, and S/N for
the training sample. As expected, parallaxes correlate very well with
$\teff$ and $\logg$, while S/N anti-correlates with the apparent
magnitude. No correlation was measured between $\mh$ (and $\alpham$)
and apparent magnitudes. We discuss the correlation matrices of the
observed sample in Sect.~\ref{kiel_diagram}. We note that
we did not use S/N as an extra information for the CNN. The S/N is
naturally encoded in the spectra, and we show in Sect.~\ref{stability_snr}
that the CNN is extremely stable across all S/N ranges.

\begin{table}
\caption{\label{label_range_training_sample}Effective range of training sample labels, parallax, and G magnitude.}
\centering
\begin{tabular}[c]{l l}
\hline
\hline
Label  & Effective range \\
\hline
$\teff$   & [ 3\,705 : 6\,395 ] K \\
$\logg$   & [ +0.58 : +4.70 ] \\
$\mh$     & [ $-$2.29 : +0.55 ] \\
$\alpham$ & [ $-$0.18 : +0.46 ] \\
$\feh$    & [ $-$2.20 : +0.54 ] \\
G         & [9.01 : 14.21] mag \\
parallax  & [0.05 : 7.00] mas \\
\hline 
\hline
\end{tabular}
\end{table}


\section{Convolutional neural network for \emph{Gaia}-RVS}\label{cnn_method}

In the present study, we adopted a hybrid CNN approach that was
first successfully applied in \citet{guiglion2020} with RAVE
spectra and \emph{Gaia} DR2 astrometry and photometry. A complete
description of the CNN can be found in \citet{guiglion2020},
\citet{Ambrosch2023}, and \citet{Nepal2023}. We built our CNN models
with the open-source deep learning library Keras \citep{chollet2015keras}
using the TENSORFLOW backend \citep{tensorflow2015}.


\begin{figure}[h!]
        \centering
        \includegraphics[width=1.0\linewidth]{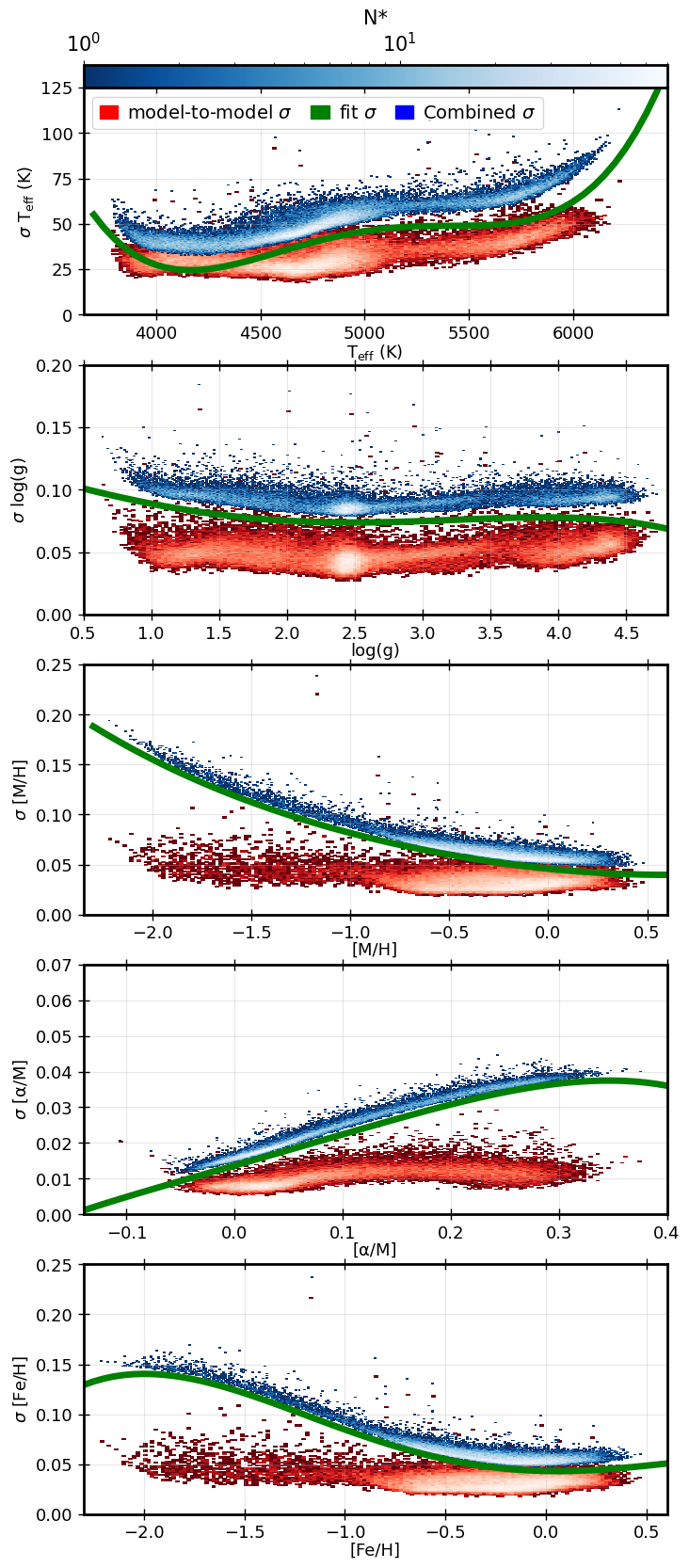}
        \caption[]{Convolutional neural network uncertainties of the training sample as a function of CNN
 output labels. The  2D density distribution in the red colourmap
 corresponds to the internal precision computed over the 28 CNN models. The
 green fit corresponds to the running dispersion computed from the residual
 of CNN-APOGEE training labels (see \figurename~\ref{fig:training_vs_gspspec}).
 The 2D histogram in the blue colourmap corresponds to the quadratic sum of the internal precision and running dispersion, and defines our overall uncertainty.}
        \label{fig:uncertainty_plot_training}
\end{figure}

\subsection{Convolutional neural network principles}\label{cnn_principles}

Following our previous works and what has been largely
adopted in the community when dealing with stellar parametrisation, 
we adopted a CNN
approach. Convolutional neural networks are well known for being sensitive to spectral features 
and learning from such features, as well as being less sensitive to
radial velocity shifts in the spectra than simple artificial
neural networks (see for instance \citealt{Nepal2023}
and references therein). The CNN allows for the building of a high-dimensional
non-linear function that translates spectra plus extra data to
stellar labels. The architecture of the CNN we employed is built
on the architecture of the CNN developed by  \citet{Nepal2023}. 
We note that \citet{Nepal2023} and \citet{Ambrosch2023}
extensively improved CNN architectures for spectroscopy compared
to \citet{guiglion2020}. We therefore refer the reader to the former two papers
for more technical details. We used keras\_tuner
\citep{omalley2019kerastuner} to further optimise the CNN architecture
and fix the model and training hyperparameters.

\begin{figure*}[h]
        \centering
        \includegraphics[width=\linewidth]{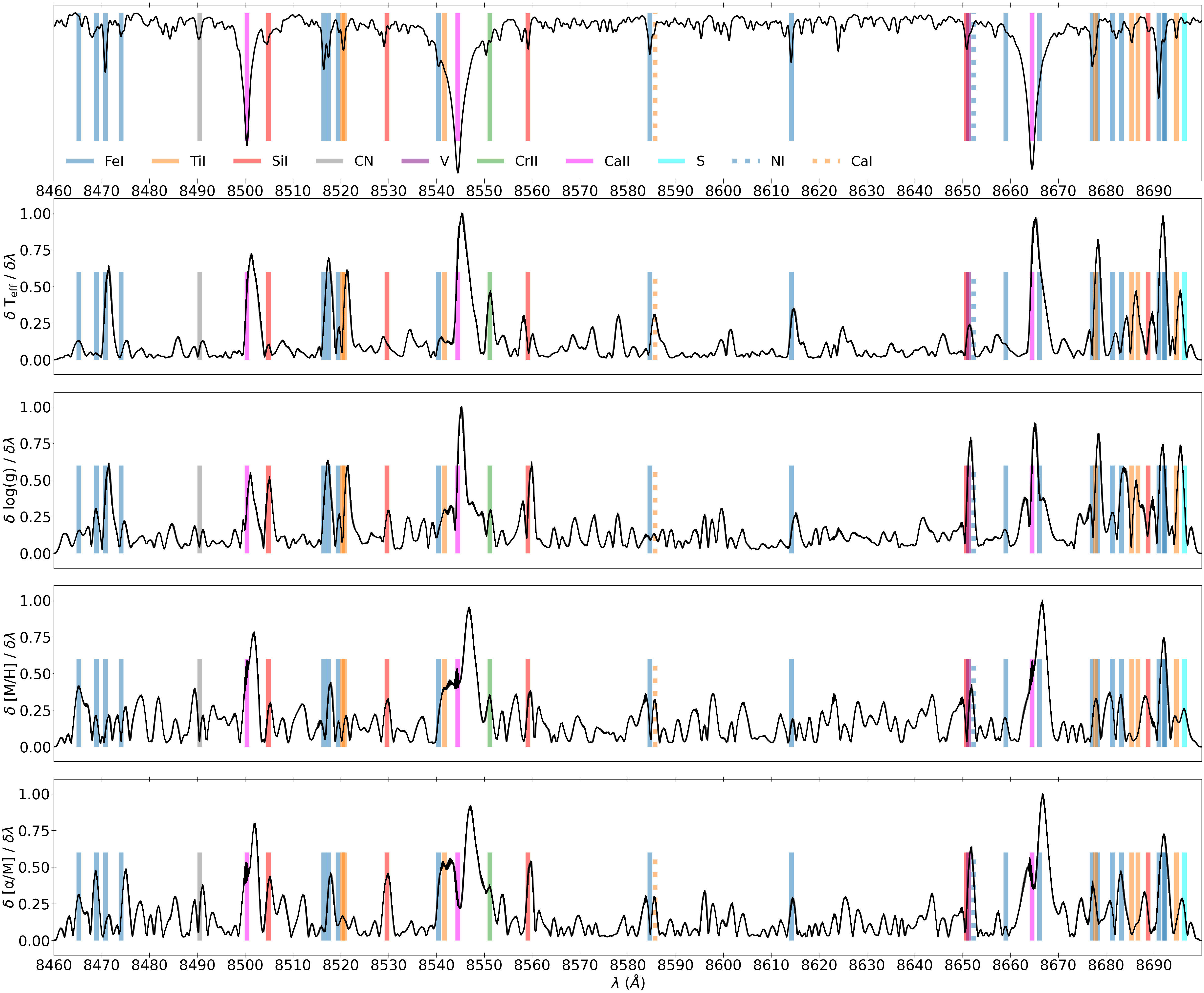}
        \caption[gradients]{\emph{Gaia} RVS spectral sensitivity maps of CNN. Top: Mean RVS spectra of training sample RC stars with
 $\mh\sim0$. In the next sub-panels, we show the mean gradients of the CNN output labels
 with respect to the input RVS pixels for $\teff$, $\logg$, $\mh$, and $\alpham$
 ($\delta \mathrm{label} / \delta \lambda$). The vertical coloured lines 
 show the location of the main RVS spectral features from which CNN learns (see Sect.~\ref{rvs_gradients}).}
        \label{fig:grad_rvs}
\end{figure*}

The hybrid CNN model developed in this paper consists of three input nodes. The core of our approach consists of \emph{Gaia}-RVS
spectra passed through a block of three 1D convolution layers (with 32,
16, and 8 filters in each convolution block, respectively)
that focus on extracting the relevant
spectral features sensitive to the stellar labels. The first convolution
block has 2\,401 input neurons, corresponding to 2\,401 pixels of
the RVS spectra (see central node in \figurename~\ref{fig:cnn_architecture}).
After extensive testing, we adopted a kernel size of 8 pixels for each convolution block,
larger kernels would not have allowed for the detection of small spectral features. We used 1D 
Max-Pooling layers (after 1D convolution blocks two and three) that help
the network focus on important features, in addition to reducing the
number of parameters to fit in the CNN. The output of the third
convolution layer was then passed
through a block of fully connected layers with 128 neurons.

The second node consists of \emph{Gaia} DR3
apparent magnitudes G, $\mathrm{G}\_{\mathrm{BP}}$, and $\mathrm{G}\_{\mathrm{RP}}$ together with the parallax
serving as four input neurons fully connected to a layer of 32 neurons
(see left node in \figurename~\ref{fig:cnn_architecture}). We adopted
LeakyRelu activation functions for the fully connected layers
\cite{xu2015empirical}, which are commonly
adopted in the community as well as in our previous works.

The third input node consists of XP coefficients passed in the
form of 110 input neurons, corresponding to the 110 coefficients, to
a fully connected layer with 64 neurons (see right node in
\figurename~\ref{fig:cnn_architecture}).

The outputs of these three fully connected layers were then concatenated
(total of 32+128+64=224 neurons), combining
the information from all three sources, and passed into three
fully connected layers with 128, 32, and 5 neurons each. The last layer
with five neurons refers to the output corresponding to the five labels, namely,
$\teff$, $\logg$, $\mh$, $\alpham,$ and $\feh$. To facilitate a faster and
more efficient convergence of the CNN to the global minimum of the loss function,
we scaled the stellar labels to values between zero and one. Additionally, we applied
the same scaling procedure to the magnitudes and parallax. In the case of XP
coefficients, the 55 BP coefficients were normalised relative to the first BP
coefficient (corresponding to 15th magnitude; \citealt{Andrae2023b}) and then
scaled between zero and one. Similarly, we performed the same scaling procedure for the 55 RP coefficients.

We adopted the ReduceLROnPlateau callbacks from Keras in order to reduce
the number of training epochs and hence the computation time. In order to prevent
overfitting and to stop the training when the loss function of the
validation set reached its minimum, we used the early-stop callbacks
with a patience of 20.

\subsection{Training an ensemble of convolutional neural networks}\label{cnn_ensemble}

As the weights and biases of a CNN model are initialised stochastically
at the beginning of each training phase, the predicted labels can
vary between different models.
The training sample is usually split randomly into a training\footnote{Throughout the paper, `training sample' refers to the whole data used
for training and cross-validation purposes; `train set' and `validation set' refer
to 75\% and 25\% of the `training sample', respectively.} set (seen by
the CNN at each training epoch) and a validation set (used at each epoch to
cross-validate with the training set and optimise the CNN weights). Usually,
the splitting of the training sample into training and validation sets is only performed once and frozen to train the
CNN. In the present study, we adopted a new approach: We split the training
sample into seven different training and validation sets with seven random states
(keeping a constant training-validation ratio of 75\%). In other
words, the CNN experienced seven representations of the training sample so that
the whole training sample would pass through the CNN. For each random state,
four models were trained, leading to 28 CNN models. The 28 trained CNN models
were used to predict labels (28 times) for the whole training sample
(44\,780 star) and the observed sample (841\,300 stars). In a given sample, 
the labels were averaged over the 28 models, while the standard deviation
provided an estimate of
the CNN's internal uncertainties. Such a deep-ensemble CNN approach allowed
for a more efficient exploration of the gradient space, which helps with generalising and reduces
the variance and bias
\citep{lee2015_deep_ensemble, Bialek2020,mudasir2021_deep_ensemble_review}.
Deep ensembles are also more efficient when training on large datasets and
improve feature selection.

The CNN models reached their minimum validation loss function typically after 80
epochs. The training time of one model was about 8 minutes on an Apple M1 Macbook
Pro laptop (with a total time of 4 hours for the 28 CNN models), and the prediction time was
$\sim0.3$ milliseconds per star. In other words, the whole observed sample of $8\times10^5$
RVS stars took about 4 minutes to compute (total time of $\sim$2 hours for 28 models).

\begin{figure}[h]
        \centering
        \includegraphics[width=1.0\linewidth]{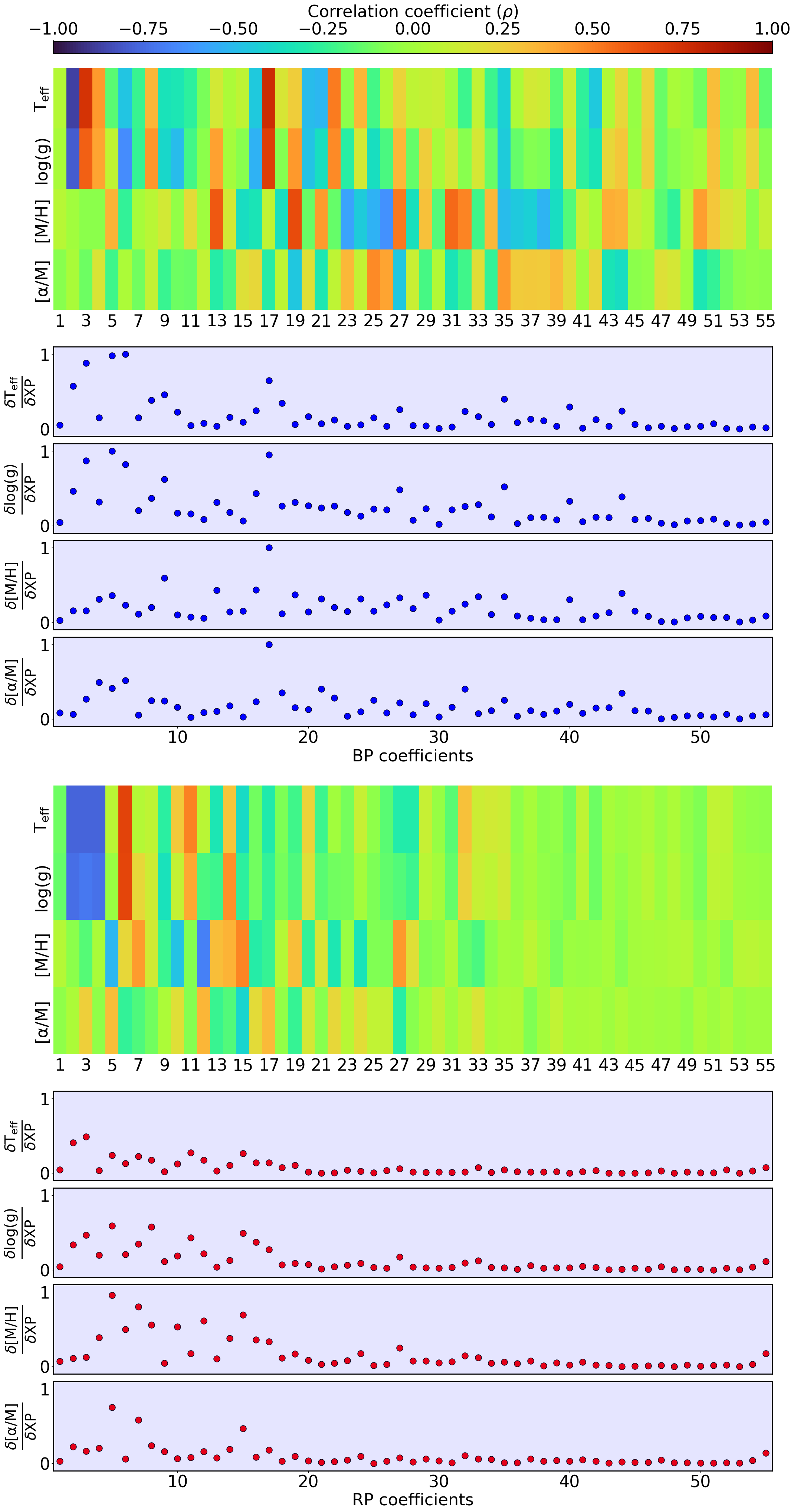}
        \caption[gradients\_XP]{\emph{Gaia} XP spectra sensitivity maps of CNN. In the top panel, we show a correlation matrix shown as a heat map between the 55
 BP XP coefficients and labels in the training sample. The colour bar shows the
 strength of the (anti-)correlation, where green in the middle (zero value) represents
 no linear relationship between the labels and XP coefficients. The second panel shows the mean
 gradients of BP XP coefficients ($\delta \mathrm{label} / \delta \mathrm{XP}$) as
 a function of the 55 BP XP coefficients for the training sample. The third and fourth
 panels depict the heat map and gradients for the RP XP coefficients, ie. the red part of XP spectra.}
        \label{fig:gradients_XP}
\end{figure}

\begin{figure*}[h]
        \centering
        \includegraphics[width=\linewidth]{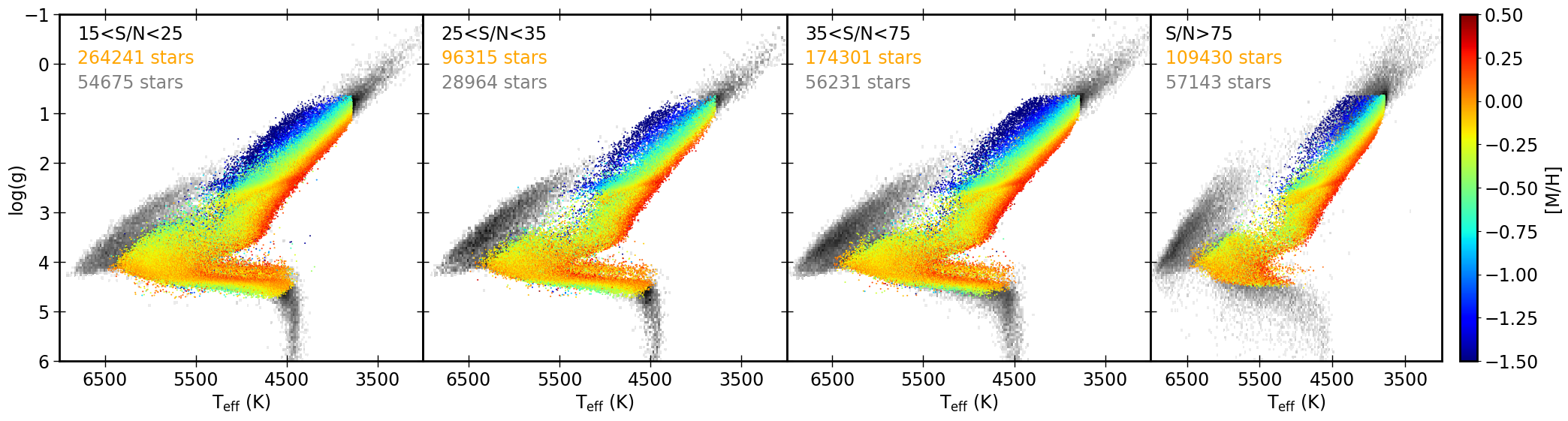}
        \caption[kiel_obs]{Kiel diagrams of 644\,287 \emph{Gaia}-RVS stars in
 bins of S/N selected within the training sample limits.
 The stars are colour-coded by metallicity. In the background, we show 2D
 histograms (in grey) of 197\,013 stars that fall outside of the training sample limits.}
        \label{fig:obs_snr_cut}
\end{figure*}

\subsection{Results of the training}\label{results_training}

In the left column of \figurename~\ref{fig:training_vs_gspspec}, we display
2D histograms of the difference between the CNN-trained labels and (input)
APOGEE labels as a function of (input) APOGEE labels for the training sample.
The dispersion between the input and output labels is 59\,K in $\teff$, 0.11\,K
in $\logg$, 0.07\,dex in $\mh$ and $\feh$, and 0.04\,dex in $\alpham$, which
is remarkable for RVS spectra. We do not show the $\feh$ results, as they are
almost identical to$\mh$ (APOGEE DR17 $\mh$ tracks $\feh$; \citealt{Abdurrouf2022}).
For the gravities, no significant systematic offset was
detected, and the red clump locus is well reproduced. On the edges
of the training sample range, we measured larger residuals (-200\,K for
$\teff>6200\,$K, -0.13 dex for $\alpham>+0.35$) due to the smaller number of
stars in these regions of the training sample. The CNN was able to measure
$\mh$ well, even if one can see a slight residual trend of about -0.04\,dex for
$\mh>-0.5\,$dex and +0.1/+0.2\,dex in the very metal-poor regime ($\mh<-2$,
due to a lack of training stars in this region of the parameter space).
Overall, the tiny mismatch
between the input and output labels tells us that the CNN is not likely to overfit
and that the small residuals are likely to come from the fact that we combine
spectra (with a different wavelength coverage and resolving power compared to
APOGEE), \emph{Gaia} DR3 photometry, parallaxes, and XP spectra.

We also show how GSP-Spec\footnote{https://doi.org/10.17876/gaia/dr.3/43}
atmospheric parameters and $\alpham$ \citep{recioblanco2023} compare to APOGEE
for the training sample stars. We adopted GSP-Spec
$\teff$ and calibrated $\logg$, $\mh$, and $\alpham$ from \citet{recioblanco2023}.
We note that the authors performed basic polynomial calibration of $\logg$
and $\mh$ using external parameters from APOGEE DR17, GALAH DR3, and RAVE DR6,
while $\alpham$ was calibrated using the local galactic abundance trend.
We applied the recommended GSP-Spec flags, setting the first 13 digits of flags\_gspspec to zero (see Table 2 from \citealt{recioblanco2023}) and resulting in 2\,606 stars in common with our training set.
We show comparison plots in the right panels of
\figurename~\ref{fig:training_vs_gspspec}.
Overall, the dispersion is two to three
times larger compared to the CNN. One can also see the presence of large outliers
in $\teff$ (>1000\,K bias) and $\logg$ (>1\,bias), likely due to the fact that
RVS spectra alone present limited resolution and spectral coverage. We
elevated the results for the RVS dataset to the level of the APOGEE survey both
in terms of precision and accuracy, significantly improving the results from GSP-Spec,
which did not use external information as we do with our hybrid CNN.
Such systematics in GSP-Spec results were recently
reported by \citet{brandner2023benchmarking}.

\begin{figure}[h]
        \centering
        \includegraphics[width=\linewidth]{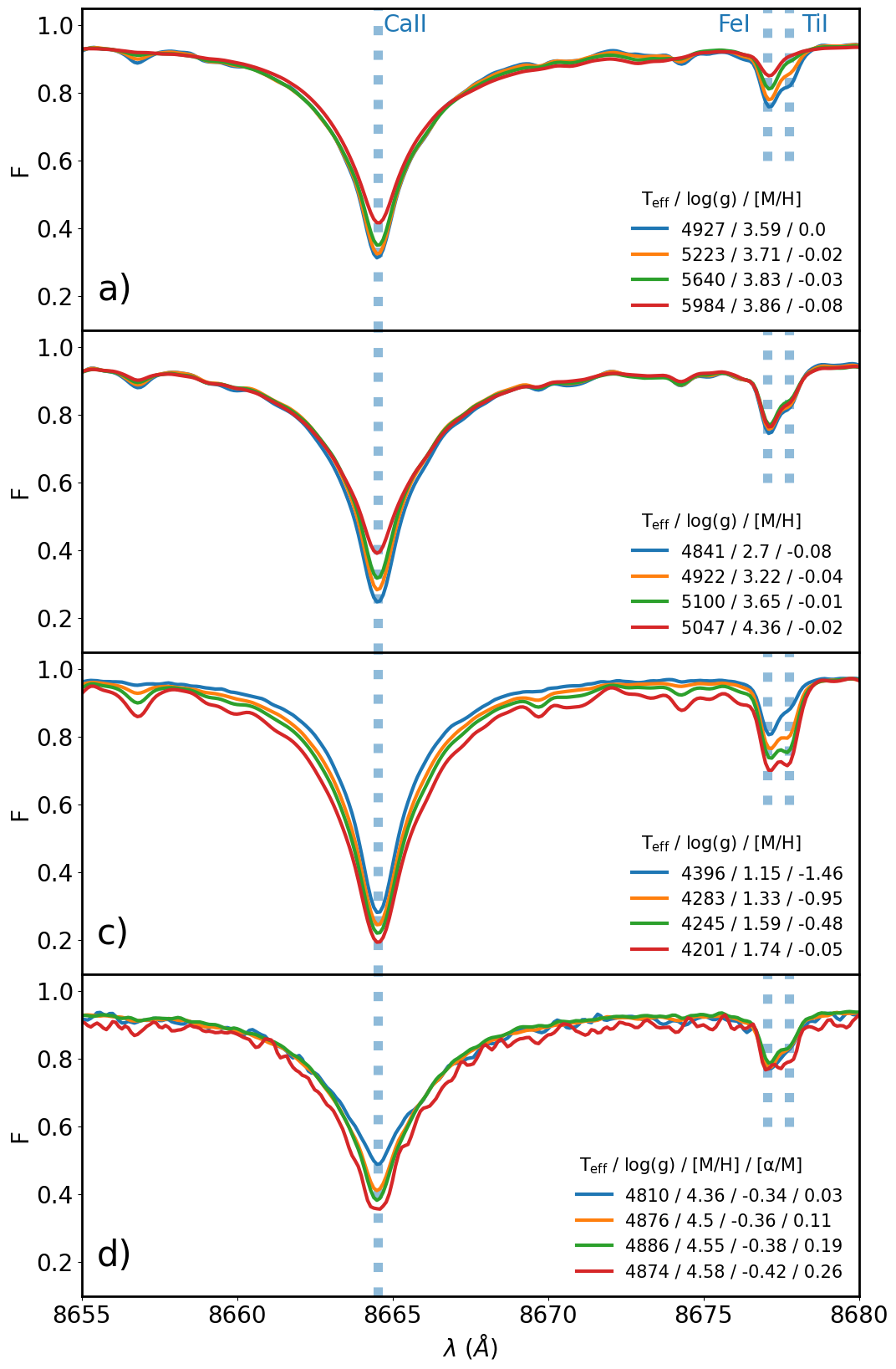}
        \caption[]{Mean RVS-observed spectra plotted in sequences of $\teff$ (panel a.), $\logg$ (panel b.), $\mh$ (panel c.), and $\alpham$ (panel d.). We identified the two main spectral features as \ion{Ca}{II} and a blend of \ion{Fe}{I} and \ion{Ti}{I}.}
        \label{fig:spectral_sequence}
\end{figure}

\subsection{Determination of training sample uncertainties}\label{errors_section}

The internal CNN uncertainty of the stellar labels is given by the standard deviation of the 28 CNN-trained models described in Sect.~\ref{cnn_ensemble}. \citet{guiglion2020} and \citet{Nepal2023} showed the past that such an internal model-to-model dispersion may
not be representative of the expected uncertainty at a given spectral resolution and may be an underestimate of the true uncertainty. This internal uncertainty
may not be representative of the precision of the CNN, as it does not reflect the
precision from the input labels. In order to provide more realistic uncertainties,
we proceeded as in \citet{Nepal2023}. Therefore, in the training sample, we computed the mean
dispersion between the APOGEE input and the CNN output labels as a function of the APOGEE
input labels. Such a dispersion gives an estimate of the precision with respect to
the training sample input labels that are considered as ground truth.

The results are shown in \figurename~\ref{fig:uncertainty_plot_training}. The internal
dispersion over the 28 CNN models (red colourmap) is on the order of 30$-$40\,K in
$\teff$, 0.05\,dex in $\logg$, 0.03/0.05 dex in $\mh$ and $\feh$, and 0.01\,dex in
$\alpham$. The green line in the figure represents the running dispersion of the difference between
the CNN output and the APOGEE input labels and shows how the training is precise compared
to the ground truth. When quadratically combining the internal dispersion and the
running dispersion, the total uncertainty significantly increases (in blue). We
note a 50\,K uncertainty in $\teff$ for the giants and 60$-$70\,K for the dwarfs.
The $\logg$ is rather constant, around 0.1\,dex. The $\mh$ and $\feh$ precision increase
from 0.06/0.08\,dex in the intermediate metallicity domain to 0.15\,dex in the
very metal-poor regime, due to the paucity of spectral features. The $\alpham$ is
extremely precise, with a typical precision on the order of 0.02/0.04\,dex. We note
that such a remarkable precision for all labels at RVS resolution is only achievable
when combining external information in the form of \emph{Gaia} DR3 photometry,
parallaxes, and XP coefficients.

\begin{figure}[h]
        \centering
        \includegraphics[width=\linewidth]{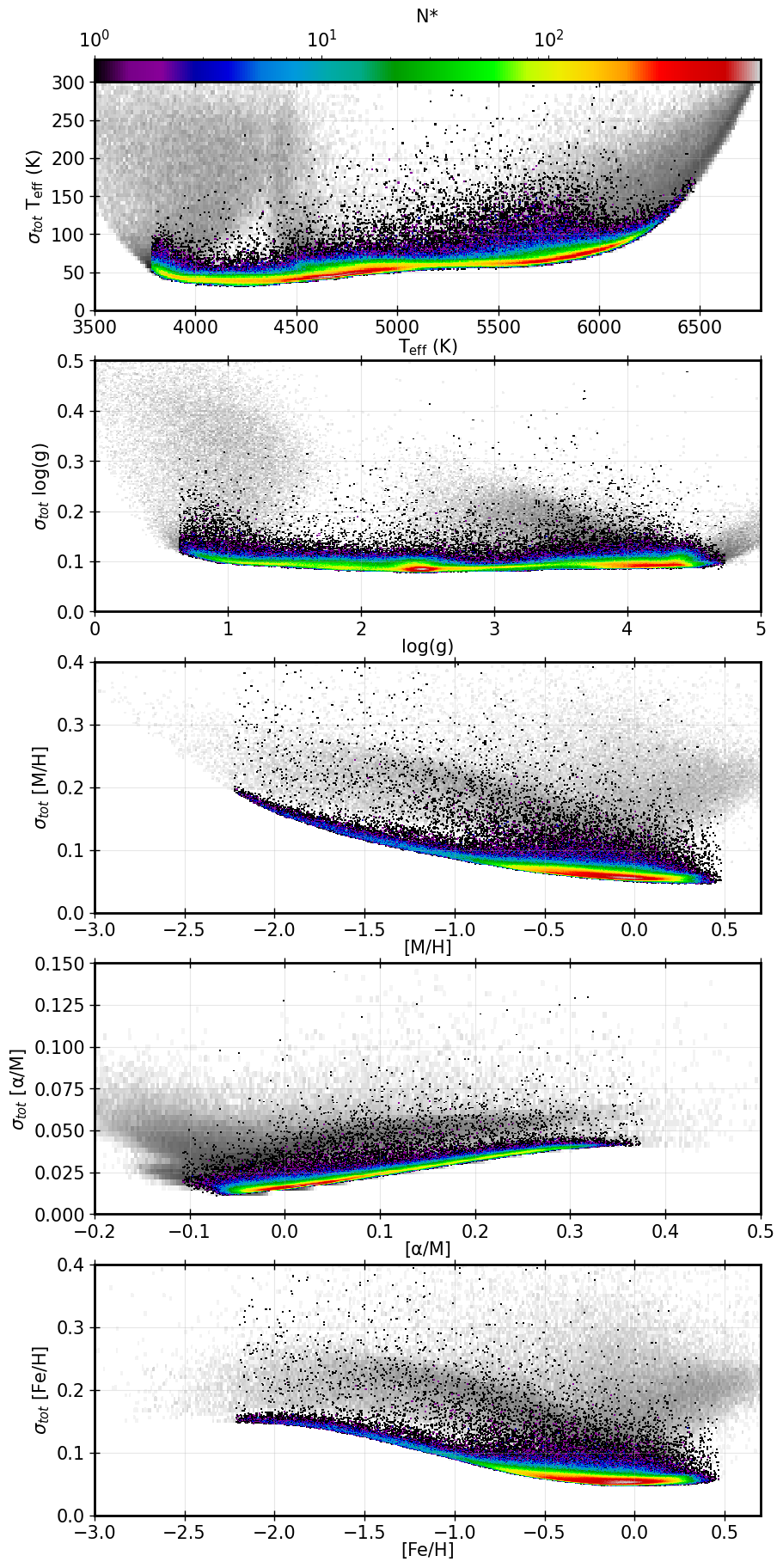}
        \caption[]{Two-dimensional density distribution of the total uncertainties for
 the 644\,287 observed sample stars within the training sample limits as a function
 of stellar labels. In the background, we show in grey the uncertainty distributions for stars outside of the training set limits (see Section~\ref{outside_ts_limits}).}
        \label{fig:uncertainty_plot}
\end{figure}

\subsection{Exploring the convolutional neural network gradients}\label{section_gradients}

We provide here a comprehensive view of the features used by the CNN during the training process. We present feature maps for RVS spectra in Sect.~\ref{rvs_gradients} and XP feature maps in Sect.~\ref{xp_gradients}.

\begin{figure*}[h]
        \centering
        \includegraphics[width=\linewidth]{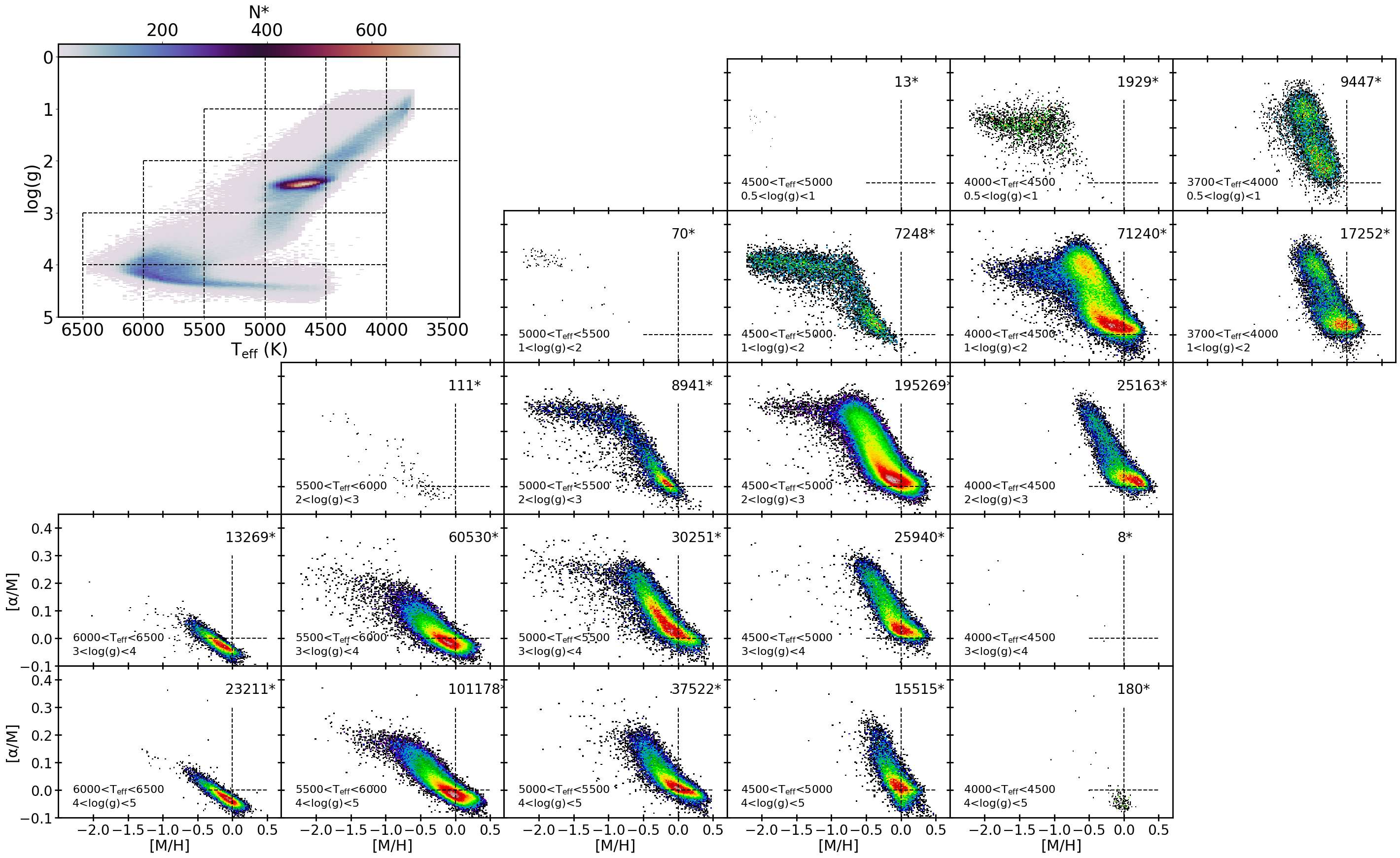}
        \caption [alpha_fe_obs]{Representation of $\alpham$ versus $\mh$ for 644\,287 \emph{Gaia}-RVS stars of the observed sample within the training sample limits.
    The sample is presented in panels
    corresponding to cuts in the effective temperature and surface
    gravity (steps of 500\,K in $\teff$ and 1 dex in $\logg$). In the
    top-left corner, we show a Kiel diagram of the sample to guide the eye.}
        \label{fig:alpha_fe_obs}
\end{figure*}

\begin{figure}[h]
        \centering
        \includegraphics[width=\linewidth]{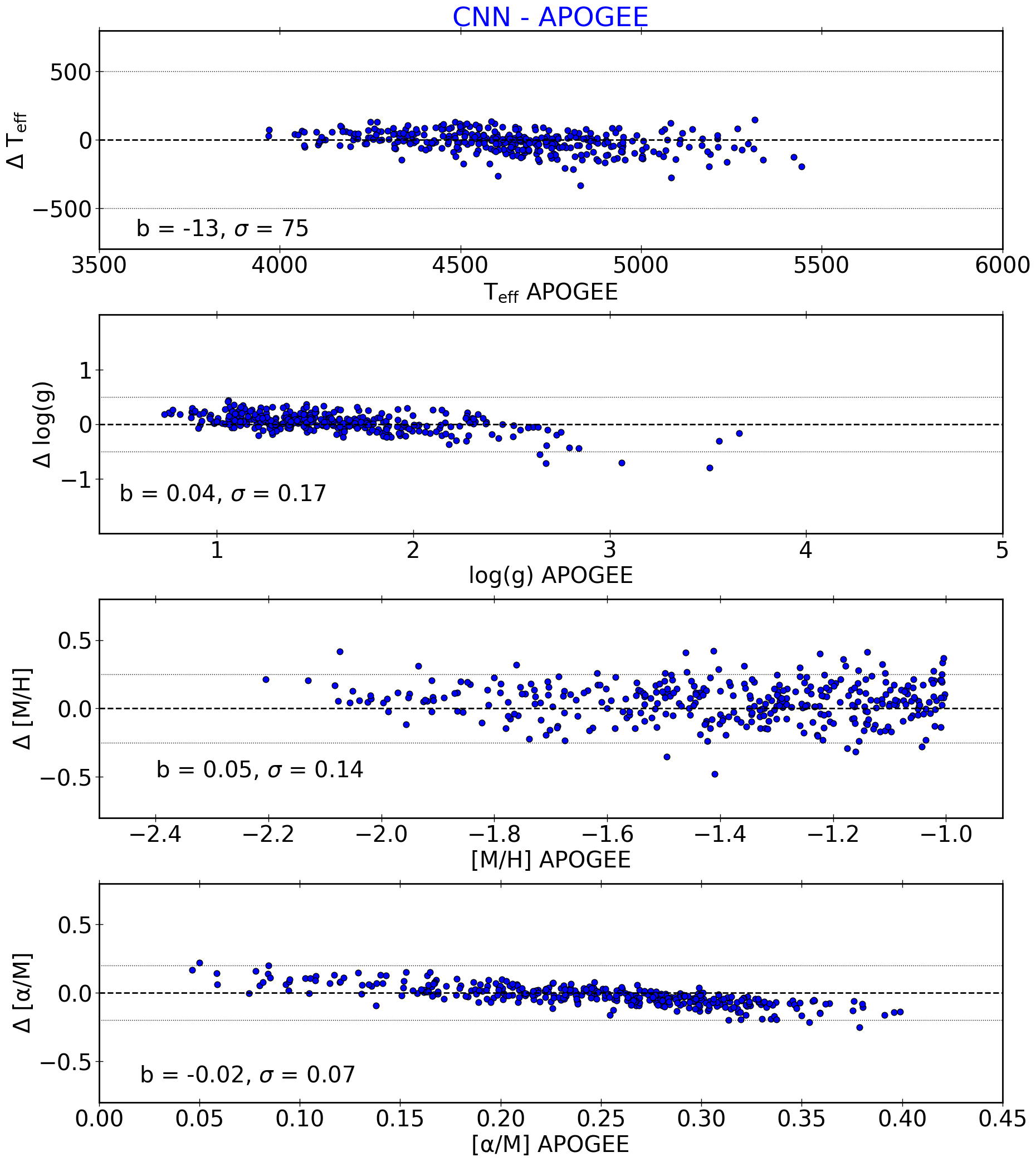}
        \caption[]{Residual between the CNN and APOGEE parameters as
 a function of APOGEE for 353 metal-poor stars ($\mh_\mathrm{APOGEE}<-1\,$dex) 
 in the observed sample in the range $15<$S/N$<25$. The black dashed line shows a null difference. 
 The mean bias (b) and dispersion ($\sigma$) of the difference is given in
 the bottom-left corner.}
        \label{fig:metal_poor_observed}
\end{figure}

\subsubsection{RVS gradients}\label{rvs_gradients}

As demonstrated in \citet{Nepal2023}, \citet{Ambrosch2023}, as well as in
\citet{fabbro2018}, CNNs are able to learn each label from specific spectral
features. We computed CNN gradients for the training sample RVS spectra
by performing partial derivatives of each of the labels with respect
to each input neuron (or pixel), namely, $\delta \mathrm{Label} / \delta \lambda$.
Such gradients provide comprehensive maps of the active spectral features
during the CNN training. The RVS gradients are shown in \figurename~\ref{fig:grad_rvs}.
We present mean gradients for the solar $\mh$ (solid line) and $\mh\sim-0.8\,$dex
(dashed line) RC stars. We present some characteristic features used by the CNN
that were taken from various literature sources \citep{boeche2011, Guiglion2018RNAAS,
Contursi2021}. Apart from being the strongest feature in the \emph{Gaia}-RVS range,
the CaII triplet is not the most prominent feature in the gradient maps.
In the blue wing of the \ion{Ca}{II} 8544 line, the \ion{Fe}{I}+\ion{Ti}{I}
blend is active for all labels. We observed that the \ion{Cr}{II} line at
8\,551\AA~is active for $\teff$ and $\logg,$ but almost no signal is present in
$\mh$ and $\alpham$ gradients. An \ion{Fe}{I} and \ion{Fe}{I}+\ion{Ti}{I} blend
at $\sim8\,520\,$\AA~is mainly active in $\teff$, $\logg$, and $\mh$. A very
interesting feature can be seen at $\lambda\sim8\,650\,$\AA: a blend of \ion{Si}{I},
\ion{V}{II,} and N. This blend is mainly active in $\logg$, $\mh$, and $\alpham$
gradients. The training was done on APOGEE labels for which we knew the
APOGEE C and N features correlate with mass \citep{Salaris2015, Martig2016}
and therefore also with APOGEE metallicity and $\alpham$. This explains why the N feature is used 
by the CNN for constraining $\logg$, $\mh$, and $\alpham$. An \ion{Fe}{I} blend
in the red wing of the \ion{Ca}{II} line at $8664\,$\AA~seems to be a very relevant
feature for the determination of the four labels. We note that most of the spectral
features used in the range 8\,570-8\,640\,\AA~are composed of \ion{Fe}{I} lines. In
the red part of the RVS domain, we observed numerous lines contributing to gradients:
an \ion{S}{I} line is active in $\teff$ and $\logg$ gradients as well as in \ion{Ti}{I}.
Such lines are also active in $\mh$ and $\alpham$, but to a lower extent. From what
we observed, the CNN learns $\alpham$ from mainly \ion{Si}{I}, \ion{Ca}{II}, \ion{and Ti}{I}
lines and from the blend of Si, V, and N. Overall, the CNN is able to learn from
distinct spectral features for a given label even if the resolution is $\sim10\,000$.
This bodes well for the future exploitation of 4MOST low-resolution surveys
\citep{chiappini2019, Helmi2019, Cioni2019}.

\subsubsection{XP gradients}\label{xp_gradients}

We investigated how the network learns from the XP coefficients. To that end, we
first computed correlations between the labels and the 110 XP coefficients. As
presented in \figurename~\ref{fig:gradients_XP}, the correlation matrix
tells us that some coefficients are more correlated to labels than others.
It is evident from the heatmap in the figure that XP spectra contain a lot of information
on the stellar atmospheric parameters $\teff$, $\logg$, $\mh$, and $\alpham$.
Interestingly, some of the XP coefficients show a good amount of correlation with the alpha
abundance as well. We then expected the CNN to learn differently from each XP
coefficient. In the same way we did for the RVS spectra, we computed gradients
for the 110 XP coefficients. We present the XP gradient
($\delta \mathrm{XP} / \delta \mathrm{label}$) as a function of XP coefficients
in the right panel of \figurename~\ref{fig:gradients_XP}.
The gradients show a lot of activity, meaning that the CNN uses and learns from
the XP coefficients for the training sample. The basis functions with high orders
in RP ($>30$) do not seem to show strong activity, meaning that the CNN may learn
most from the lower-order coefficients. In case of BP, information is present even
down to coefficient 45. The gradients are fairly consistent with the correlations
observed between XP and labels. Such plots confirm that XP coefficients are extremely
rich in information (see \citealt{Andrae2023b, Zhang2023}) and can provide
additional constraints when measuring $\teff$, $\logg$, $\mh$, $\feh$, and $\alpham$.


\section{Predicting labels for the observed sample}\label{prediction_observed_sample}

Using the 28 CNN models, we predicted the atmospheric parameters
$\teff$, $\logg$, $\mh$ as well as $\feh$ and $\alpham$ ratios
for 841\,300 \emph{Gaia}-RVS stars, that is, the previously defined
observed sample together with their uncertainties
as detailed in Sect.~\ref{errors_section}. Among these 841\,300
stars, we have 644\,287 stars within the training sample limits
as defined in \tablename~\ref{label_range_training_sample} and Sect~\ref{tsne_concept}.

\subsection{Kiel diagrams of the observed sample}\label{kiel_diagram}

In \figurename~\ref{fig:correlation_matrix_training_set}, we show a
correlation matrix of labels, photometry, parallaxes, and S/N of the RVS spectra
classified as within the training sample limits (middle column). Correlations
are consistent with those from the training sample (left column), as expected.
On the other hand, when drawing a correlation matrix for the rest of the observed sample
(outside the training sample limits), we did not see any strong correlation of
$\teff$ with parallaxes. In fact, G and $\mathrm{G}\_{\mathrm{RP}}$ are anti-correlated
with $\teff$. Such a behaviour is discussed in more detail in
Sect.~\ref{explore_labels_lim} and is mainly due to the presence of underrepresented spectral
types in the training sample.

In \figurename~\ref{fig:obs_snr_cut}, we show Kiel diagrams of the
observed sample in bins of S/N (15-25, 25-35, 35-75, and $\ge75$)
colour-coded with $\mh$ (644\,287 stars). We observed a very
consistent Kiel diagram with a clear metallicity
sequence in the giant branch. The CNN also does a good job of parameterising the red clump.
We note that the CNN catalogue contains 10\,718 RVS stars with $\mh<-1$.
We also observed a secondary cool-dwarf sequence, which is very similar to what has been observed
in RAVE DR6 \citep{steinmetz2020a} and in which case results from the presence of
binary stars. Indeed, these stars in the regime $4\,600<\teff<5100\,$K and
$4<\logg<4.2$ show a very large \emph{Gaia} DR3 RUWE ($\sim4$),
suggesting a poor astrometric solution likely due to binarity. We
further discuss the stability of the CNN with radial velocity errors in
Appendix~\ref{vrad_test}. 

We also present Kiel diagrams for the rest of the RVS sample (197\,013 stars) in grey (stars outside the training sample limits). Such stars are discussed in more detail in Section~\ref{outside_ts_limits}. In Appendix~\ref{only_rvs}, we provide more detail regarding the CNN application using only RVS spectra (i.e. no photometry, parallaxes, or XP data).

In order to show that the CNN is able to properly parameterise the RVS spectra, we show in \figurename~\ref{fig:spectral_sequence} the mean spectra from the observed sample in sequences of $\teff$, $\logg$, $\mh$, and $\alpham$ as derived by the CNN. Panel a of the figure shows typical turn-off stars with solar $\mh$, ranging from $\sim4\,900$ to 6\,000\,K. As expected, cooler stars present shallower spectral lines. Next, panel b shows a $\logg$ sequence from 2.7 to 4.4 dex for stars around 5\,000\,K and solar $\mh$. In the same fashion as $\teff$, cooler stars present shallower CaII feature, while the \ion{Fe}{I} and \ion{Ti}{I} blend do not show strong sensitivity to gravity. Panel c shows a metallicity sequence from -1.5 to solar $\mh$ for cool giants. Overall, the more metal-poor stars suffer from the weakening of spectral lines, as expected. Panel d shows a $\alpham$ sequence for cool dwarfs from Solar to +0.26. Similar to $\mh$, shallower lines result from lowering the overall $\alpham$ ratio. The \ion{Ti}{I} component of the blend seems more sensitive to $\alpham$ enrichment. Such diagnostic plots show that the CNN properly determines stellar labels and propagates the knowledge from the training sample labels.

\subsection{Uncertainties of the observed sample}

The uncertainties were derived by quadratically combining the dispersion from
the 28 CNN models together with the fit of the running dispersion from the
training sample (polynomial curve in \figurename~\ref{fig:uncertainty_plot_training}).
In \figurename~\ref{fig:uncertainty_plot}, we present the total uncertainty of
the 768\,793 stars within the training sample limits. The bulk of the sample shows
very similar uncertainty distributions to those in the training sample (see
\figurename~\ref{fig:uncertainty_plot_training}). Stars with larger uncertainties
are also present. For instance, cool super giants show uncertainties on the order
of 70-300\,K in $\teff$, 0.15-0.5 dex in $\logg$, and 0.1-0.3 dex in $\mh$
(roughly 19\,000 stars). We note that we provide to the community both the model-to-model
dispersion and the overall combined uncertainty (see \tablename~\ref{catalogue}).

\subsection{The $\alpham\,$ versus\,$ \mh$ distributions of the observed sample}\label{alpha_fe_obs}

We explore in this section the abundance pattern of $\alpham\,$ versus $\,\mh$ of the observed
sample in the different regions of the Kiel diagram when selecting stars
within the training sample limits (644\,287 stars). $\alpham\,$ versus $\,\mh$ patterns are presented in \figurename~\ref{fig:alpha_fe_obs} in bins of 
500\,K in $\teff$ and 1\,dex in $\logg$. As expected, we probed the
low-$\alpham$ regime preferentially in the dwarf regime, as such objects
are likely located closer to the Sun. We started probing the high-$\alpham$
sequence when moving to lower $\logg$, and we started populating the metal-poor
tail of the sample. In the region $4\,000<\teff<4\,500\,$K and $1<\logg<2$,
we clearly observed a bimodality in $\alpham$, as expected (e.g. 
\citealt{hayden2015, queiroz2020}). Such a feature was not visible
using RAVE data \citep{guiglion2020}. We discuss the
bimodality in more detail in Sect.~\ref{science_verification}. 

\subsection{Precision and accuracy of metal-poor stars in the observed sample}\label{section_metal_poor_observed}

We investigated the precision and accuracy of metal-poor stars 
present in the observed sample that fall within the training sample limits.
As a reference, we used APOGEE DR17 labels
and focused only on the stars with $\mh_\mathrm{APOGEE}<-1\,$. To demonstrate the
robustness of the CNN in the low S/N regime, we selected stars with $15<$S/N$<25$. We 
note that GSP-Spec do not provide results with good quality flags\_gspspec for these stars. In
\figurename~\ref{fig:metal_poor_observed}, we compare the CNN
labels to APOGEE DR17 for 353 RVS metal-poor stars of the observed sample.
The effective temperature and
surface gravity show no significant bias, with a dispersion of 75\,K and
0.17 dex, respectively. The overall $\mh$ also shows no significant bias,
with a dispersion of 0.14\,dex. The bottom panel
of \figurename~\ref{fig:metal_poor_observed} shows an $\alpham$ dispersion below 0.07\,dex and a tiny residual that is a
function of $\alpham$ and is consistent with the training sample
(see Sect.~\ref{results_training}).

Taken together, these results demonstrate that the CNN is able to provide robust parameterisation of metal-poor stars down to $\mh=-2.3\,$dex at $15<$S/N$<25$. Notably, this is the S/N regime
where standard spectroscopy struggles to provide precise and accurate
measurements for such types of stars.

\subsection{Stability of the convolutional neural network in the low signal-to-noise ratio regime}\label{stability_snr}

\begin{figure}[h]
        \centering
        \includegraphics[width=\linewidth]{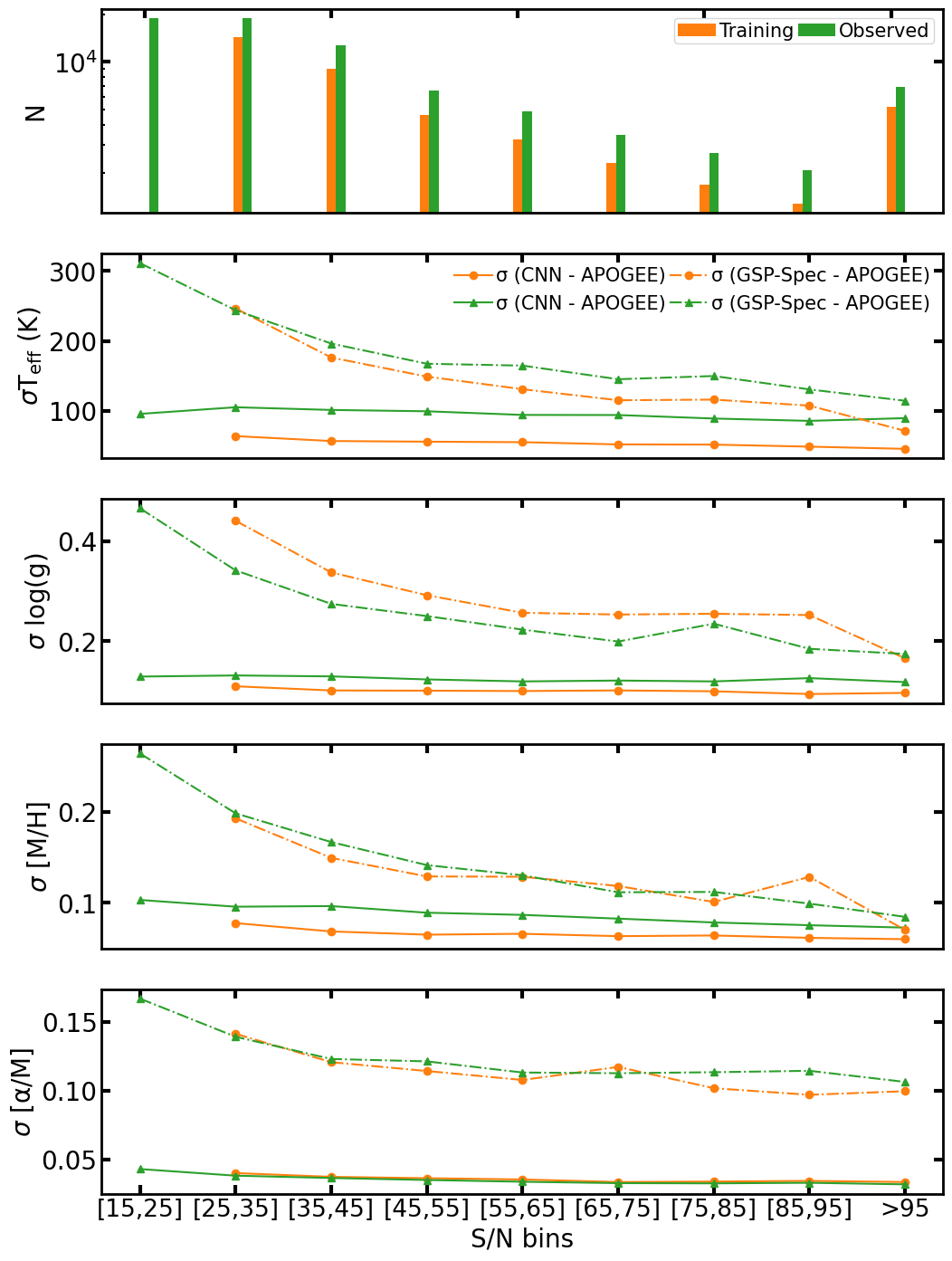}
        \caption{Precision computed as the standard deviation of the CNN minus
 APOGEE (solid lines) and calibrated GSP-Spec minus APOGEE (dashed lines)
 as a function of S/N bins for 41\,623 stars of the training sample
 (orange) and 76\,996 stars of the observed sample (green, within the
 training sample limits).}
        \label{fig:snr_stability}
\end{figure}

The \emph{Gaia} DR3 RVS sample contains a significant fraction of
low S/N stars: 38\% of the observed sample spectra range from 15
to 25 in S/N. In this section, we investigate how the CNN precision varies
with the S/N. Thus, we computed the standard deviation between the CNN
labels and APOGEE DR17 for stars within the training sample limits
in the observed
sample. We computed the same quantities in the training sample for reference.
In \figurename~\ref{fig:snr_stability}, we show how the precision behaves
as a function of the S/N. For the training sample (in orange, solid lines),
the CNN precision is extremely stable and constant with respect to the S/N for our four
labels ($\teff$, $\logg$, $\mh,$ and $\alpham$). As a comparison, the
precision of the GSP-Spec with respect to APOGEE is two to five times worse
(in orange, dashed lines), showing decreasing precision with a decreasing
S/N. We observed the same behaviours in the observed sample (in green).
The CNN precision is constant as a function of the S/N in the observed sample 
as well, which is not the case for GSP-SPec, again due to the fact that the CNN combines 
spectroscopy, photometry, and astrometry. Such plots show how
the CNN is able to efficiently deal with the noise
in the data compared to standard spectroscopic methods, and it is able to
extract high-quality labels in low S/N spectra. Such an advantage of the CNN will
be key for the next data releases of \emph{Gaia}, for instance, for extracting
information from individual epoch spectra at very low S/N, where epoch
spectra will be released at intrinsically lower S/N than the presently available
time-averaged spectra. In Appendix~\ref{vrad_test}, we present additional CNN sensitivity tests with respect to radial velocity uncertainties.

\section{How to ensure that the convolutional neural network labels are within the physical limits of the training set}\label{outside_ts_limits}

Machine learning algorithms, such as the CNN, are extremely
proficient at learning from spectral features present in a training sample spectra. Nevertheless, some spectra of the observed sample may not share common features with the training sample. Hence, parameterising such types of spectra could lead to systematics in the determined labels. To understand how reliable the CNN labels are, we present in the next sections a classification method based on t-SNE, and we explore in Section~\ref{explore_labels_lim} the labels of stars outside of the training limits.

\subsection{Using the t-SNE method to understand the limitations of convolutional neural network labels}\label{tsne_concept}

\begin{figure*}[h]
        \centering
        \includegraphics[width=1.0\textwidth]{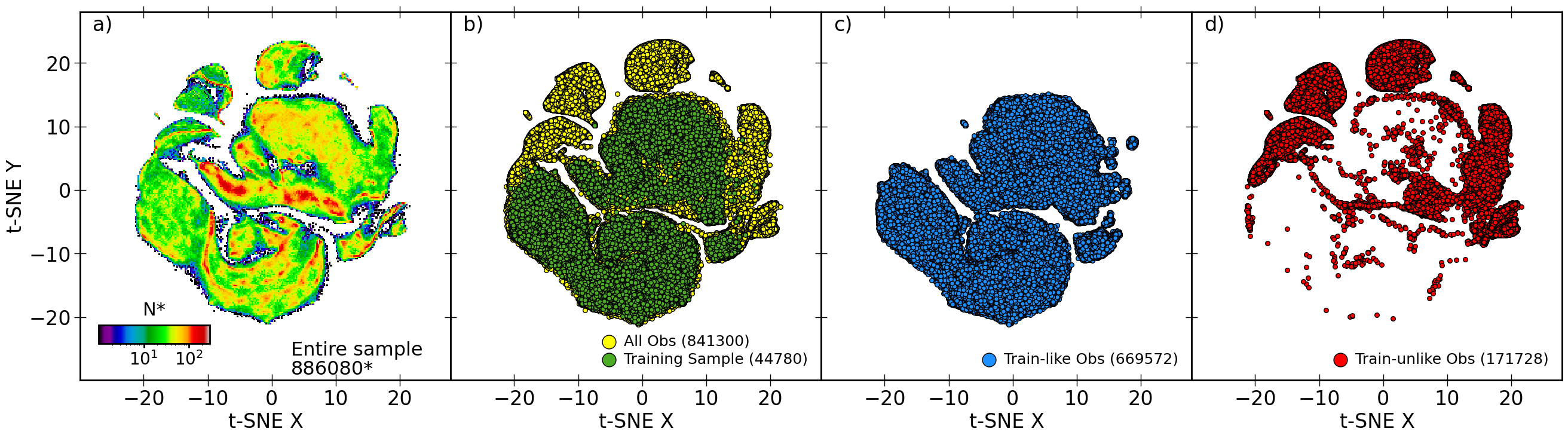}
        \caption{Classification of \emph{Gaia} RVS spectra with t-SNE. Panel a): t-SNE maps with perplexity=50 of the entire RVS sample consisting of 886\,080 spectra plotted in a 2D-histogram manner. Panel b): Same map but split into training (green) and observed (yellow) samples. Panel c): Spectra of the observed sample identified as similar to that of the training sample. Panel d): Spectra of the observed sample identified as not being similar to that of the training sample. See Sect.~\ref{outside_ts_limits} for more details.}
        \label{fig:tsne_concept}
\end{figure*}

In order to supplement the reliability of our CNN results, we performed a classification of our observed sample spectra into `training-like' and `training-unlike' spectra, in the same fashion as in \citet{Ambrosch2023}. For such a task, we used t-SNE, which is a dimensionality reduction technique \citep{van_der_maaten_2008}. From an N-dimensional dataset (i.e. full spectra), t-SNE will provide a 2D map where each point corresponds to a data sample, and points close to each other in such a map then share similar spectral features. We concatenated the training sample (44\,780 spectra) and the observed sample (841\,300 spectra) into a main sample of spectra (886\,080 spectra), in all composed of 2\,401 pixels. We produced four different t-SNE maps with perplexity=[30, 50, 75, 100]. We emphasise that this hyperparameter is equivalent to the number of neighbours for a given data point (see \citealt{van_der_maaten_2008} for more details). We give an example of a t-SNE map with perplexity=50 for the RVS sample in \figurename~\ref{fig:tsne_concept}. Panel a of the figure shows a density map of the 886\,080 spectra, colour-coded by the number of spectra per bin. In panel b, we display the same map but highlight the training sample spectra in green and the observed sample spectra in yellow. One can clearly see that in some regions, only yellow points are visible, meaning that such observed spectra do not share the same spectral features as the training sample spectra. We computed geometric distances between each point of the training sample and the observed sample in the t-SNE map for the four different perplexities. We selected observed spectra similar to the training sample simultaneously for the four learning rates (an observed spectrum must be similar to a training sample spectrum in each of the four computed t-SNE maps). A similar method was already used in \citet{Ambrosch2023} for selecting training-like observed spectra. In panel c, we show 669\,572 RVS spectra that are training-like, while panel d shows 171\,728 training-unlike RVS spectra. In Appendix~\ref{tsne_appendix}, we show similar plots for the four classifications in \figurename~\ref{fig:tsne} and plots for the classification with perplexity=75
colour-coded with $\teff$, $\logg$, and $\mh$ in \figurename~\ref{fig:tsne_teff_logg_mh}.

\subsection{Defining a robust flag to isolate spurious convolutional neural network labels}\label{flag_definition}

Thanks to the t-SNE classification, we were able to isolate the CNN labels that may suffer
from systematics. Additionally, \emph{Gaia} $G$, parallaxes, and labels of the observed sample outside of the physical limits of the training sample (as described in
Table~\ref{label_range_training_sample}) can suffer from systematics. We provide an
eight-digit integer flag in which each digit corresponds to one of the labels (in the order $\teff$,
$\logg$, $\mh$, $\alpham$, and $\feh$)  as well as the \emph{Gaia} G magnitude, parallax,
and t-SNe classification. For instance, `00000000' means that all labels are within the
training sample limits and within the G magnitude and parallax of the training sample and that
the t-SNE classification considered this spectrum to be similar to the training set.
In contrast, a star with `10000000' would indicate that the $\teff$ derived by the CNN is outside of the training 
sample limits and should be taken with caution. We note that the flag we provide can
be used to search  for peculiar stars or non-FGK objects in the RVS sample.

\begin{figure*}[h]
        \centering
        \includegraphics[width=1.0\linewidth]{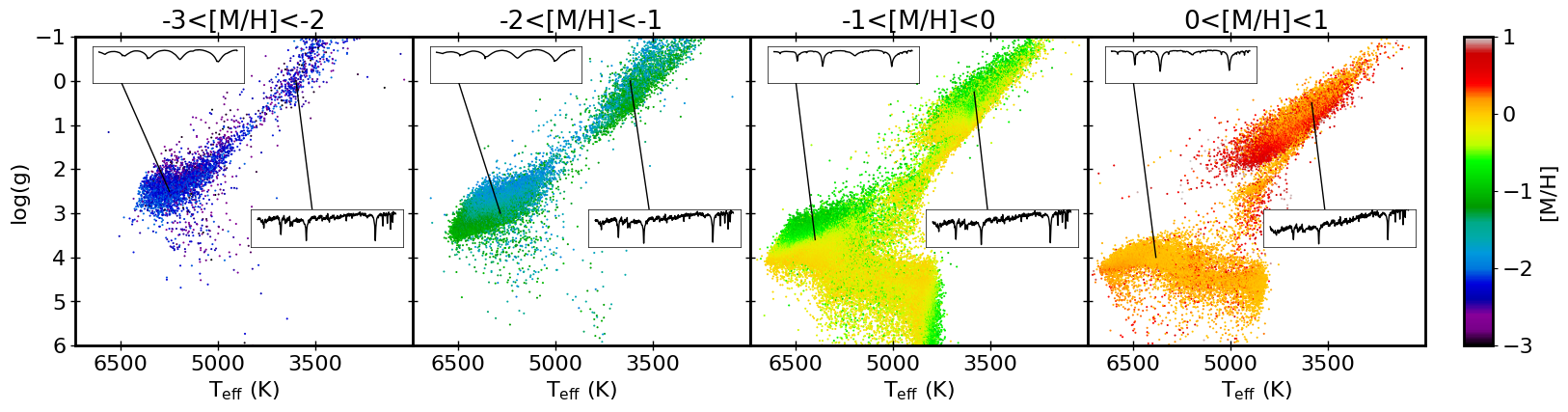}
        \caption{RVS stars outside of the training sample limits. The Kiel diagram representations (197\,013 stars) are in bins of $\mh$ and colour-coded as a function of $\mh$. We show typical RVS spectra of different regions of the Kiel diagram.}
        \label{fig:kiel_outside}
\end{figure*}

\subsection{Exploring the labels outside of the training sample limits}\label{explore_labels_lim}

In this section,
we present the CNN labels of the 197\,013 observed sample stars outside the limits presented
in \tablename~\ref{label_range_training_sample} and classified by t-SNE as training-unlike
(i.e. flag$\ne$00000000). The sample consists of 197\,013
stars, and Kiel diagrams are presented in \figurename~\ref{fig:kiel_outside}. The sample seems
to cover a large range of $\mh$ from -3 to +1.

The typical spectrum of the giant branch is presented in the bottom-right corner of each panel in \figurename~\ref{fig:kiel_outside}.
It shows very strong TiO bands that increase with metallicity. As a result, the CNN interprets the strong TiO bands as metal-rich features. Labels for these stars are unlikely to be accurate. Such TiO bands indicate that such stars are M giants (confirmed by their \emph{Gaia} DR3 spectraltype\_esphs). Such stars were also observed
in RAVE \citep{matijevic2012}. There are no such stars in the training
set.

In the top-left corner of each panel of \figurename~\ref{fig:kiel_outside}, we show spectra of the hot stars. 
The spectra show very strong Hydrogen Paschen lines, in fact indicating very hot stars.
The strength of the Paschen lines decrease with increasing metallicity, indicating that
the CNN understands the Paschen lines as being metal-poor lines. These
spectra are classified as OB stars when checking the spectraltype\_esphs
from \emph{Gaia} DR3. The CNN labels are also unlikely to be accurate.

\begin{figure}[h]
        \centering
        \includegraphics[width=1.0\linewidth]{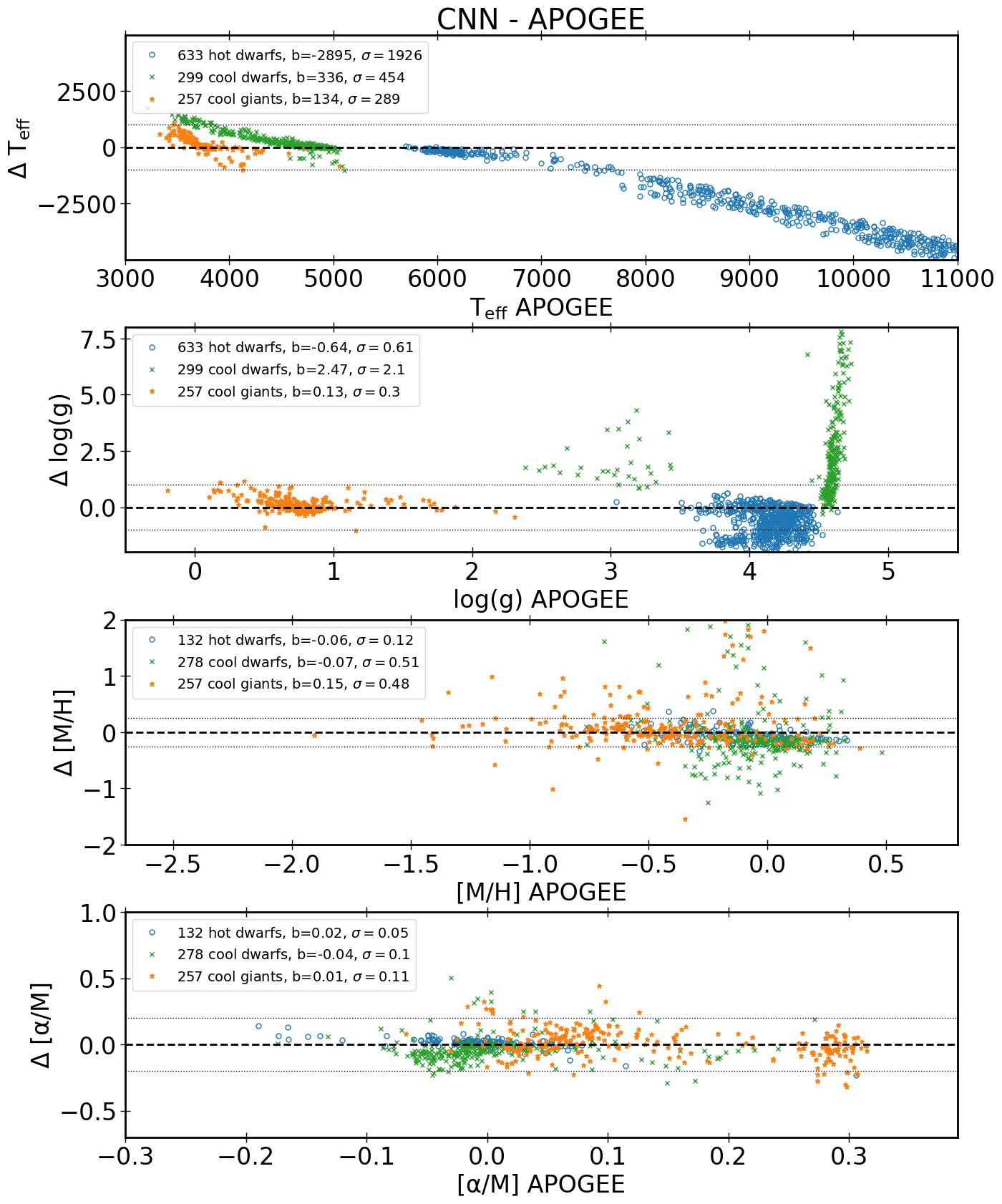}        
        \caption{Stars outside of the training set limits. The residual of
 the CNN minus APOGEE for hot dwarfs, cool giants, and cool dwarfs were selected from
 \figurename~\ref{fig:kiel_outside} (see text for more details).}
        \label{fig:outside_ts_comp}
\end{figure}

With \figurename~\ref{fig:outside_ts_comp},
we investigated in more detail three regions of the Kiel diagram that present
labels for both the CNN and APOGEE DR17:
hot dwarfs ($5\,700 \le \teff \le 7\,000$, $2 \le \logg \le 4.5$; blue dots),
cool dwarfs ($4\,000 \le \teff \le 5\,000$, $\logg \ge 4.5$; green crosses), and
cool giants ($2\,000 \le \teff \le 4\,800$, $-2 \le \logg \le 2$; orange stars). Regarding the hot dwarfs, we clearly observed that for $\teff>7\,000\,$K, there is a large discrepancy
between the CNN labels and APOGEE directly accountable for the large Paschen features in the RVS spectra and caused by the fact that the training sample does not contain such targets. The gravity suffers from large systematics as well, while $\mh$ and $\alpham$ match rather well within 0.12 and 0.05 dex, respectively.

Concerning the cool dwarfs, the main issue comes from a poorly parameterised $\logg$, as such 
objects are very nearby with parallaxes larger than 7 mas. In the current training sample,
we have few nearby cool dwarfs. Hence, the CNN gravities for such objects strongly suffer from systematics.

Finally, we observed that the cool giants show discrepant $\teff$ up to +1000\,K
compared to APOGEE, with metallicity systematics up to +1 dex. Surprisingly, the $\alpham$ agrees within 0.1\,dex
between APOGEE and the CNN.

\subsection{The catalogue of RVS labels with convolutional neural network}

We present our catalogue of atmospheric parameters ($\teff$, $\logg$
and $\mh$) along with chemical abundances ($\feh$, $\alpham$) 
for $886\,080$ \emph{Gaia}-RVS stars (summarised in \tablename~\ref{catalogue}).
We provide two sources of uncertainties, model-to-model uncertainties
(e$\teff$, e$\logg$, e$\mh$, e$\feh$, and e$\alpham$) and overall combined
uncertainties ($\sigma \teff$, $\sigma \logg$, $\sigma \mh$, $\sigma \feh$,
and $\sigma \alpham$), as well as the eight-digit flag described in
Section~\ref{flag_definition}. The data table is publicly available with the AIP
\emph{Gaia} archive\footnote{https://gaia.aip.de/metadata/gaiadr3\_contrib/cnn\_gaia\_rvs\_catalog/}. The CNN Python code can be provided upon reasonable request.
In order to use the CNN catalogue of labels, we recommend using the eight
digits to identify the stars within the training sample limits, that is, the stars
with the best CNN parameterisation. To select the CNN labels within the training set limits,
we recommend adopting flag\_boundary="00000000".

\begin{table}
\caption{\label{catalogue}Atmospheric parameters, chemical abundance ratios,
uncertainties, and boundary flag of the publicly available online catalogue of
$886\,080$ \emph{Gaia}-RVS stars.}
\resizebox{0.49\textwidth}{!}{
\centering
\begin{tabular}[c]{l l l l l}
\hline
\hline
Col  & Format & Units & Label  &  Explanations \\
\hline
1  & char  & -    & sourceid     & \emph{Gaia} Source ID                 \\
2  & float & K    & teff         & Effective temperature                 \\
3  & float & K    & eteff        & Model-to-model dispersion of $\teff$  \\
4  & float & K    & sigma\_teff   & Overall dispersion of $\teff$         \\
5  & float & \cms & logg         & Surface gravity                       \\
6  & float & \cms & elogg        & Model-to-model dispersion of $\logg$  \\
7  & float & \cms & sigma\_logg   & Overall dispersion of $\logg$         \\
8  & float & dex  & mh           & Overall metallicity                   \\
9  & float & dex  & emh          & Model-to-model dispersion of $\mh$     \\
10  & float & dex  & sigma\_mh     & Overall dispersion of $\mh$            \\
11  & float & dex  & feh          & $\feh$ ratio                           \\
12  & float & dex  & efeh         & Model-to-model dispersion of $\feh$    \\
13  & float & dex  & sigma\_feh    & Overall dispersion of $\feh$           \\
14  & float & dex  & alpham       & $\alpham$ ratio                        \\
15  & float & dex  & ealpham      & Model-to-model dispersion of $\alpham$ \\
16  & float & dex  & sigma\_alpham & Overall dispersion of $\alpham$        \\
17  & int   & -    & flag\_boundary & Boundary flag composed of 8 digits      \\
\hline 
\hline
\end{tabular}}
\end{table}


\section{Validation of convolutional neural network labels}\label{validation_section}

We validate here our CNN methodology. We using external datasets in the form of asteroseismic data, parameters from GSP-Phot, and GALAH data.

\subsection{Comparison of convolutional neural network labels with GSP-Phot}

In this section, we compare the CNN labels of the observed sample with atmospheric parameters from GSP-Phot \citep{Andrae2023a}. We note that \emph{Gaia} DR3 provided the community with spectro-photometric atmospheric parameters using parallaxes, stellar magnitude, and BP and RP coefficients.

\begin{figure*}[h]
        \centering
        \includegraphics[width=\linewidth]{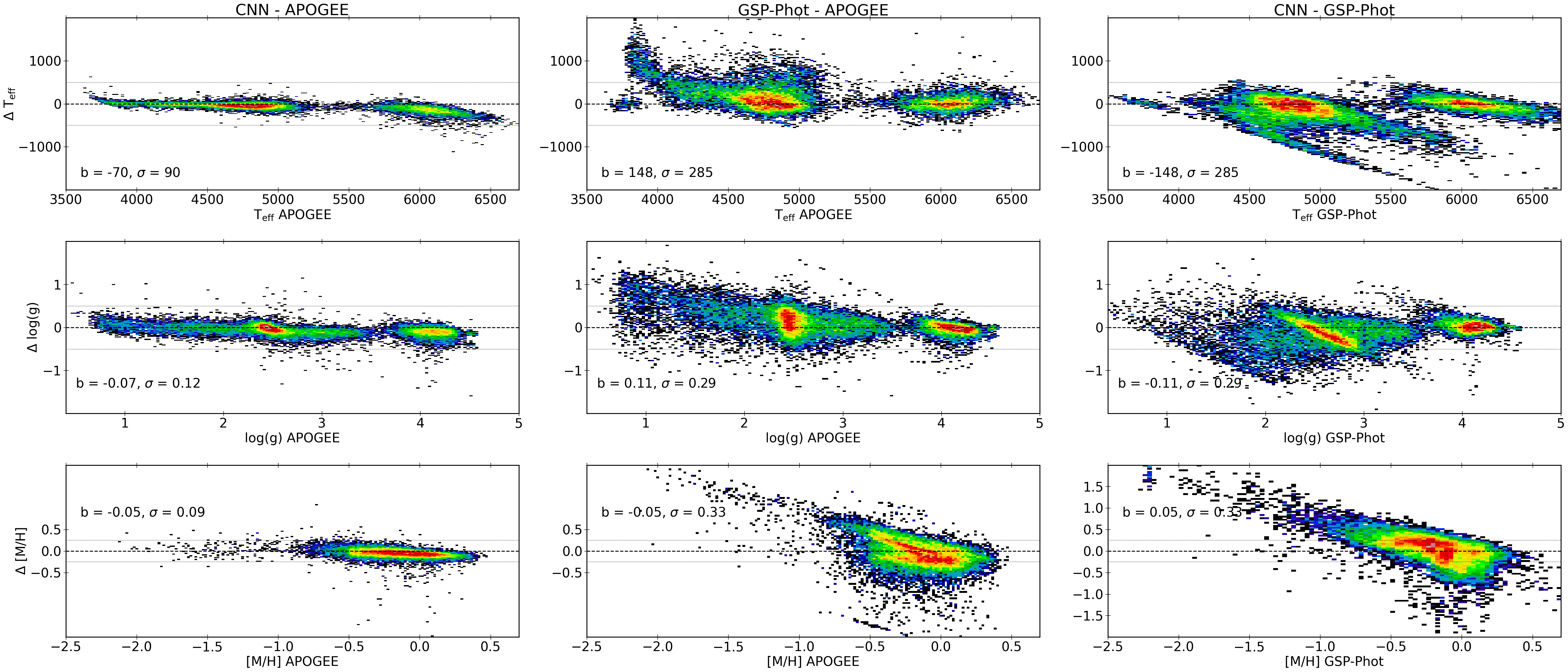}
        \caption{Atmospheric parameter comparison between CNN and APOGEE (left column), GSP-Phot and APOGEE (middle column), and GSP-Phot and CNN (right column) for
 33\,120 stars of the observed sample.}
        \label{fig:CNN_vs_GSP_Phot}
\end{figure*}

In \figurename~\ref{fig:CNN_vs_GSP_Phot}, we present comparisons between the CNN and GSP-Phot\footnote{We note that we did not apply the metallicity calibration relation proposed by \citet{Andrae2023a}.} with respect to APOGEE, as APOGEE was used as training labels. We required that the CNN labels be within the training set limits, resulting in having 33\,120 labels in common between APOGEE, the CNN, and GSP-Phot.
The left columns of \figurename~\ref{fig:CNN_vs_GSP_Phot} show a comparison of the CNN labels to APOGEE, and a very similar behaviour as seen in \figurename~\ref{fig:training_vs_gspspec} can be observed. When comparing GSP-Phot with APOGEE (middle column of \figurename~\ref{fig:CNN_vs_GSP_Phot}), a larger overall dispersion can be measured, two to three times larger than when comparing the CNN to APOGEE. There are significant systematics for $\teff<4\,500\,$K and $\logg<2$. We also noticed a double sequence in the $\mh$ residual for $\mh>-0.8$. The most striking feature is the large residual trend for $\mh<-0.8$, with a difference larger than 0.5\,dex. The third panel of \figurename~\ref{fig:training_vs_gspspec} presents comparisons between the CNN and GSP-Phot. Overall, the dwarfs compare quite well, while the giants show rather large discrepancies in both $\teff$ and $\logg$, consistent with what is shown in the left and middle panels of \figurename~\ref{fig:training_vs_gspspec}. The large discrepancy in the parameters between CNN and GSP-Phot can be explained by the fact that the CNN combines RVS spectra, astrometry, photometry, and BP and RP coefficients and that it trains on labels from the high-resolution APOGEE survey.

\subsection{Validation of surface gravities with asteroseismic data}\label{Validation_seismo}

To test CNN accuracy and precision in surface gravity, we compared
the CNN $\logg$ with the precise $\logg$ from asteroseismology. Asteroseismology
relies on stellar oscillations and is widely used by spectroscopic
surveys for validation or calibration purposes, such as in RAVE
\citep{Valentini2017}, \emph{Gaia}-ESO \citep{Worley2020}, and APOGEE
\citep{Anders2017, Pinsonneault2018, Miglio2021}. For stars with solar-like
oscillations, as well as red giants, the frequency at maximum oscillation
power ($\nu_{max}$ ) is used for determining $\logg_{\rm seismo}$
using only the additional parameter $\teff$
\citep{Brown1991, Kjeldsen1995, Chaplin2013}.
We adopted the most recent version of the K2 Galactic Archaeology Program
(K2 GAP) for campaigns C1-C8 and C10-C18 from \citet{Zinn2022}. The authors
provide asteroseismic parameters for 19\,000 red giant stars. We have 164
stars in common with our training sample and 589 stars in common with our
observed set (within the training sample limits). Each star also has GSP-Spec
calibrated $\logg$ for further comparison, with the 13 first flags\_gspec equal to zero.
We computed $\logg_{\rm seismo}$
from Zinn's $\nu_{max}$  and $\teff$ using Equation. 3 from
\citet{Valentini2019}, assuming $\nu_{max,\bigodot}=3090\,\mu\mathrm{Hz}$
and $\teffsun=5\,777$\,K \citep{Huber2011}.

\begin{figure*}[h]
        \centering
        \includegraphics[width=\linewidth]{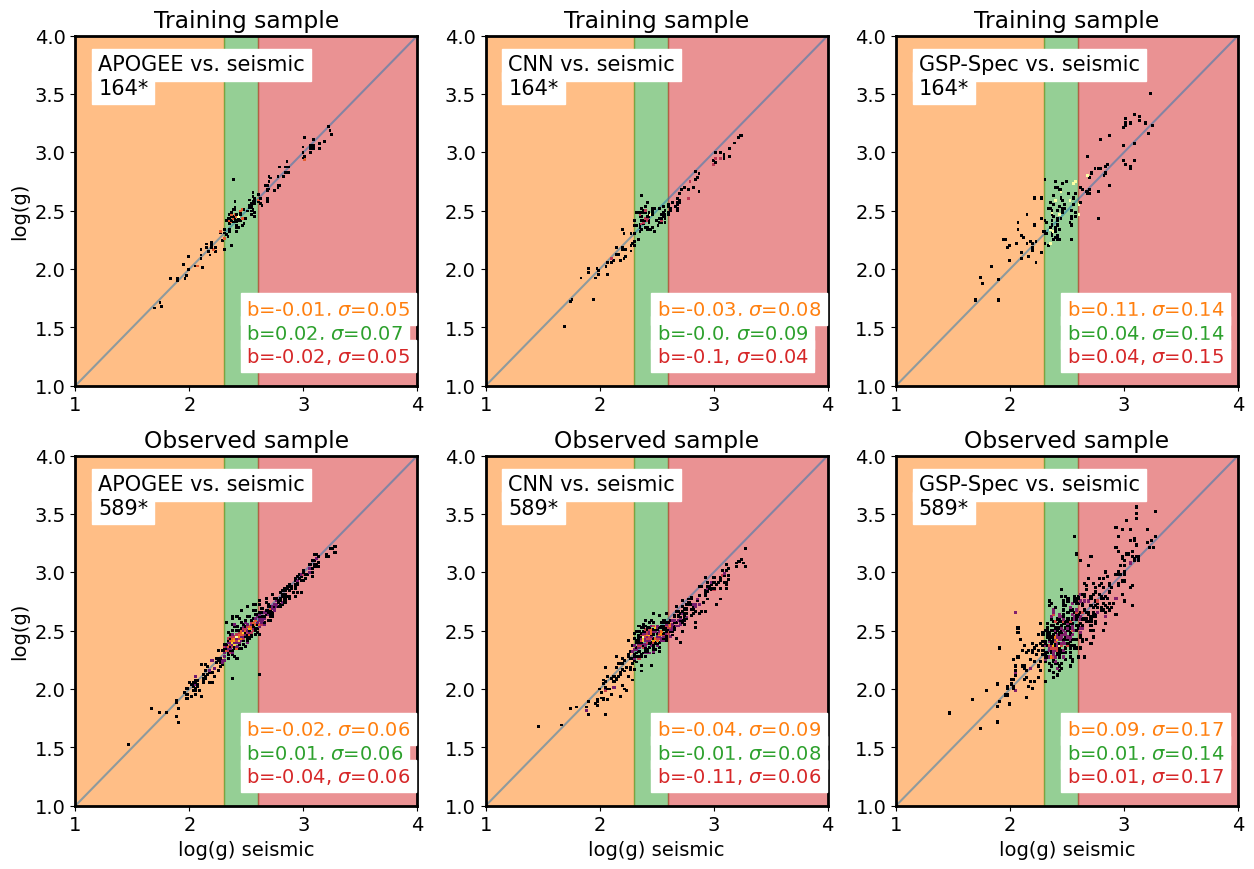}
        \caption{One-to-one comparisons of surface gravities $\logg$ from APOGEE
 DR17 (left column), CNN (middle column), and calibrated GSP-Spec (right column) with respect to
 seismic $\logg$ calculated based on $\nu_{max}$ and $\teff$ from
 \citet{Zinn2022}. The top row shows stars of the training sample, while the
 bottom row shows targets of the observed sample. We computed the mean bias
 (b) and mean dispersion ($\sigma$) for three ranges of gravity: $\logg<2.3$
 (orange), $2.3<\logg<2.6$ (green), and $\logg<2.6$ (red).}
        \label{fig:seismic_comp}
\end{figure*}

In \figurename~\ref{fig:seismic_comp}, we present a one-to-one comparison between the asteroseismic log(g)s and APOGEE DR17,
CNN, and GSP-Spec surface gravities in the training and observed
samples. Firstly, the APOGEE and seismic $\logg$ compare very well in both the training (used as input
labels) and observed stars. No significant bias was measured, while the dispersion
ranges from 0.05 to 0.07, depending on the $\logg$ range. This absence of bias is consistent
with the fact that APOGEE calibrated the spectroscopic $\logg$ with respect to
the seismic ones \citep{Abdurrouf2022}. Secondly, the CNN and seismic $\logg$ also compare very well, with no significant bias for
$\logg<2.6$. For $\logg>2.6$, we measured a small bias on the order of 0.1 dex
in both training and observed stars, likely due to the fact that we combined
spectra, photometry, parallaxes, and XP coefficients. In the observed sample,
the dispersion ranges from 0.04 to 0.09 dex, which is remarkable.
The red clump
(in green) shows no apparent bias as well as a dispersion below 0.1 dex. We
note that the $\logg$ residual between APOGEE and the CNN shows no trend with respect
to the CNN $\teff$ or $\mh$. The CNN is then capable of conserving the APOGEE calibration. Finally, we compared the GSP-Spec-calibrated $\logg$ with respect to
the seismic $\logg$. We observed that the calibrated GSP-Spec gravities match quite well the
seismic $\logg$, with a dispersion on the order of 0.15, while biases range from 0.10 to 0.11.

Such comparisons once again show the remarkable performances of the CNN, which
combines several datasets for improved measurements compared to pure spectroscopic
labels (as derived by GSP-Spec). The CNN is able to transfer and preserve the 
properties of the training set, which in the present case refers to the seismic calibration of APOGEE labels.
(We refer the reader to Appendix~\ref{only_rvs} for a CNN application using RVS
spectra only.)

\subsection{Comparison between convolutional neural network and GALAH DR3}\label{galah_comp}

\begin{figure}[h]
        \centering
        \includegraphics[width=1.0\linewidth]{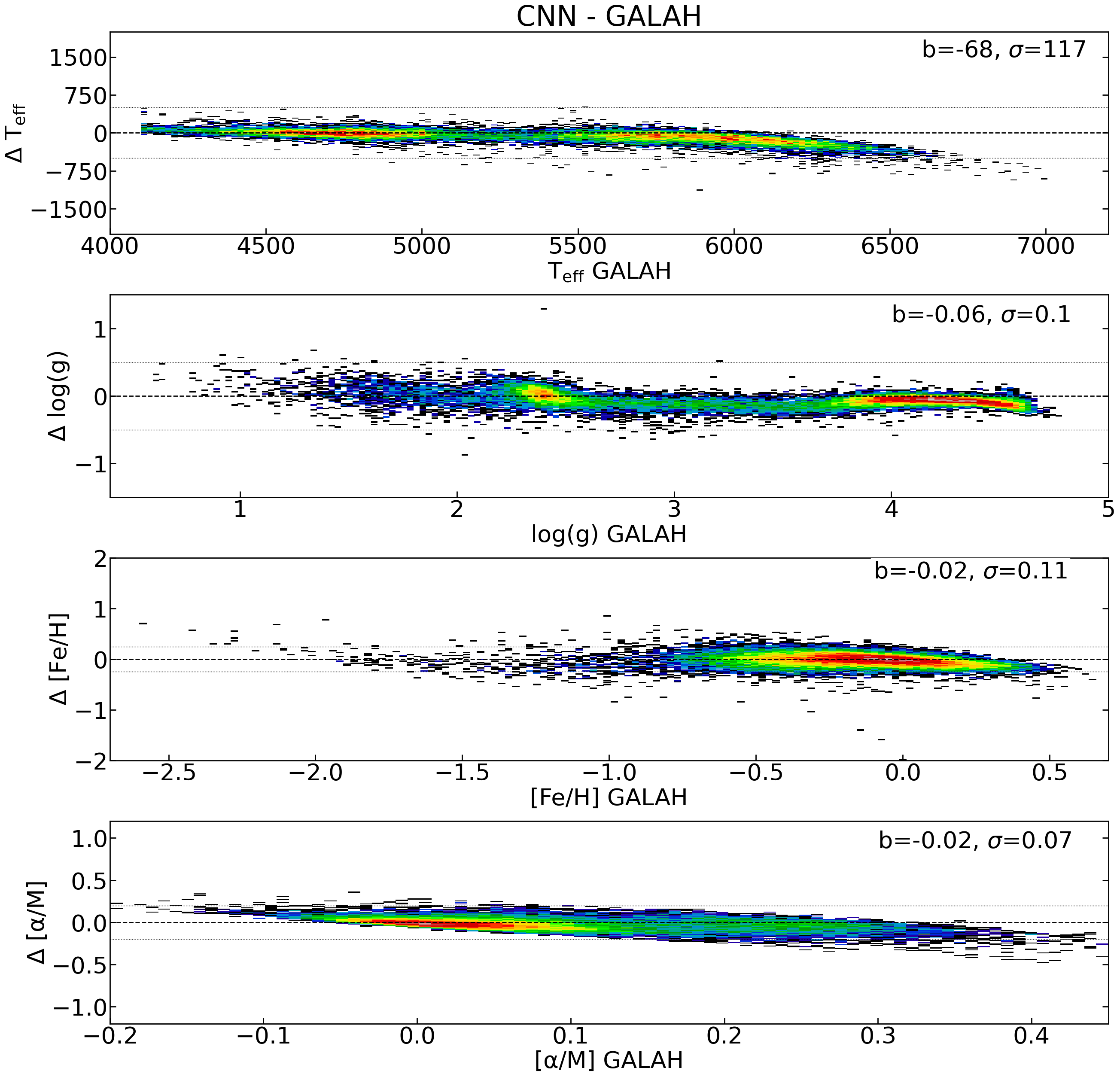}
        \caption{Differences between CNN minus GALAH as a function of GALAH
 for $\teff$, $\logg$, $\feh$, and  $\alpham$ for 24\,803 stars of the observed
 set with $15\le\snr\le25$ and within the training set limits.}
        \label{fig:galah_comparison}
\end{figure}

We previously assessed CNN performances with respect to APOGEE
(which could reflect our training sample) and GSP-Spec (which uses the
same spectra as the  CNN). In order to have a fully independent comparison,
we adopted the third data release (DR3) of the high-resolution
(R $\sim28\,000$) optical spectroscopic survey GALAH \citep{Buder2021}.
GALAH DR3  used the Spectroscopy Made Easy (SME; \citealt{valenti1996})
spectral fitting code to derive atmospheric parameters and chemical
abundances for 588\,571 stars.  We adopted GALAH quality flags according
to the recommendations presented in the best practices for using GALAH
DR3,\footnote{\url{https://www.galah-survey.org/dr3/using_the_data/}}
including removing stars flagged to have peculiarities in their
stellar parameters and iron and alpha abundances, namely, flag\_guess = 0,
flag\_sp = 0, flag\_fe\_h = 0, flag\_alpha\_fe = 0, and we made a S/N cut
with snr\_c3\_iraf > 40 per pixel. Regarding CNN labels, we only required
that labels be within the training sample limits. For completeness, we also
compared GALAH to calibrated GSP-Spec parameters. Following the above
selection, the sample consisted of 24\,803 stars in common between the CNN
and GALAH DR3 and with labels within the training sample limits and $15\le\snr\le25$.

In \figurename~\ref{fig:galah_comparison}, we present label
differences in the form CNN\,-\,GALAH as a function of GALAH.
We observed that $\teff$ and $\logg$ present no strong systematics,
apart from hot dwarf stars ($\teff>6\,500$K) with a significant bias of
>300K (total of 200 stars). For gravities, $\logg$ shows a weak residual trend with
a dispersion on the order of 0.1. We note that such a dispersion increases to 0.17 
for $\logg<2.2$. The CNN $\mh$ shows remarkable agreement with GALAH, with a dispersion below 0.1
for $\mh>-1$, and it reaches 0.19\,dex for $\mh<-1$. In addition, the CNN $\alpham$ matches GALAH well, 
with a dispersion well below 0.1. We note that the systematic trends measured between
the CNN and GALAH are also visible in the input APOGEE DR17 labels. 
In this section, we have demonstrated that the CNN parameters in the range $15\le\snr\le25$
based on spectroscopy, astrometry, and photometry are very consistent with high-resolution
spectroscopic GALAH parameters.

\subsection{Comparison of convolutional neural network iron content with open clusters from GALAH}

\begin{figure}[h]
        \centering
        \includegraphics[width=1.0\linewidth]{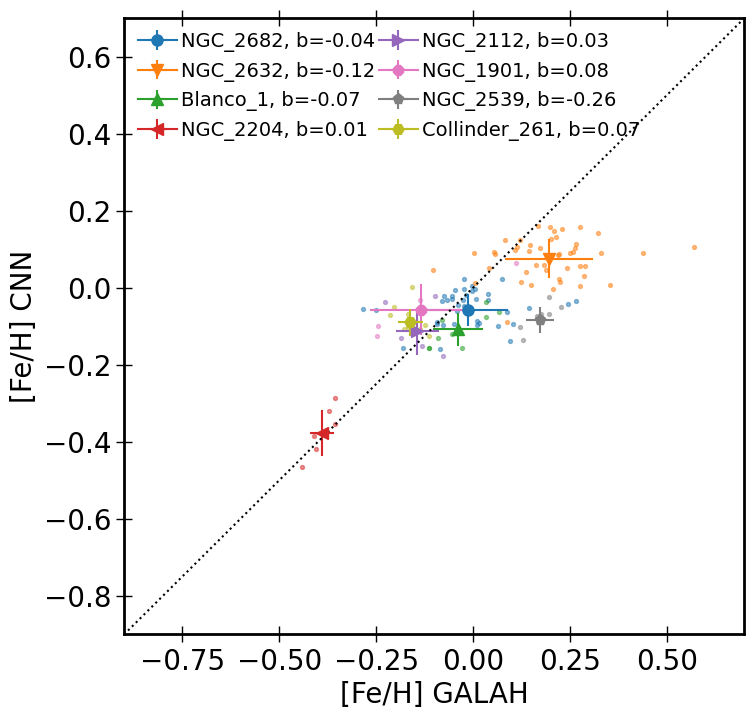}
        \caption{Comparison of CNN $\feh$ with respect to GALAH DR3 for 10 open clusters. The
 clusters are listed in the top-left corner. We also computed a mean $\feh$ and the associated error bar (computed as standard deviation). The mean bias is written next to the cluster name.}
        \label{fig:CNN_vs_clusters}
\end{figure}

We compare here the CNN $\feh$ predictions for open cluster stars to those from  GALAH DR3. We used a list of known clusters within the \emph{Gaia}-RVS clusters from
\citet{CantatGaudin2020} updated to DR3 (Cantat-Gaudin, private communication).
There are nine clusters with more than three members ranging from -0.4 to +0.25 in $\feh$. We present a comparison plot in \figurename~\ref{fig:CNN_vs_clusters}. Overall, the CNN $\feh$ agrees very well with GALAH. However, two clusters present rather large systematics. For instance, NGC 2539 shows a mean difference of 0.26 dex between the CNN and GALAH. A recent study by \citet{Casamiquela2021} found this cluster to be slightly sub-solar ($\feh=-0.012$), which is consistent with the CNN data. The second cluster, NGC 2632, shows a rather large dispersion of 0.12 dex for GALAH $\feh$, while the CNN finds an internal dispersion of 0.05 dex. The mean $\feh$ abundance is 0.08 for the CNN and 0.19 for GALAH, while \citet{Casamiquela2021} reported 0.12 as the mean $\feh$. Our comparison plot allowed us to show the robustness of the CNN $\feh$ ratios.

\section{The $\alpham-\mh$ bimodality traced by \emph{Gaia}-RVS}\label{science_verification}

The abundance patterns of $\alpha$-elements (such as magnesium and oxygen)
have been studied and characterised in the Milky Way disc, bulge, and
halo for more than two decades in the solar neighbourhood (e.g.
\citealt{Fuhrmann1998, Pompeia2002, adibekyan_2011, fuhrmann_2011,
mikolaitis_2014, guiglion_2015}) and toward the inner and outer disc
thanks to large-scale spectroscopic surveys (e.g.
\citealt{Anders2014, hayden2015, buder2019, queiroz2020}). A strong
debate has animated the Galactic Archaeology community regarding the mechanisms responsible for the  bimodality 
measure in this abundance space (see for instance \citealt{Chiappini1997, 
Schonrich2009, Haywood2013, minchev2013, Grand2018, spitoni2019, Buck2020,
Khoperskov2021, Agertz2021}). So far, the 
bimodality has been clearly characterised at high resolution, even though 
hints of such a bimodality have been detected by low-resolution and intermediate-resolution surveys, such as SEGUE \citep{lee2011}, LAMOST \citep{Xiang2019}, 
and RAVE \citep{guiglion2020}. We have shown in the previous
sections that our CNN methodology provides precise and accurate chemical
information. In \figurename~\ref{alpha_fe_obs}, we presented that giant
stars show a bimodality in the $\alpham-\mh$ plane, and
this is the first time that such a clear bimodality has been seen
in \emph{Gaia}-RVS spectra, which are characterised by both limited
resolution and spectral coverage. Our finding is fully consistent
with previous work: a low $\alpham$ sequence ($<+0.15$\,dex) ranging from
$\mh\sim-0.6$ to $\mh\sim+0.2$\,dex together with a high $\alpham$
($>+0.15$\,dex) sequence ranging from $\mh\sim-1$ to $\mh\sim-0.2$\,dex. We
note that in the literature, the shape and zero-point of the bimodality
depend on the type of stars used and the type of $\alpha$-elements studied.

To trace the spatial variations of the bimodality, we calculated the positions and
velocities in the galactocentric rest-frame using available astrometric
solutions~(sky positions and proper motions) and radial velocities from
\emph{Gaia} DR3 \citep{gaiadr3_survey_properties} and assuming
distances computed with the StarHorse Bayesian method \citep{Queiroz2023}
and the current CNN labels (more details in Nepal et al. in prep). We also
assumed an in-plane
distance of the Sun from the Galactic centre of $8.19$~kpc \citep{Abuter2018}, 
a velocity of the Local Standard of Rest of 240~\kms \citep{Reid2014}, and a
peculiar velocity of the Sun with respect to the local standard of rest, $U_\odot = 11.1$\,\kms,
$v_\odot =12.24$\,\kms, $W_\odot=7.25$\,\kms \citep{Schonrich2010}.

\begin{figure*}[h]
        \centering
        \includegraphics[width=\linewidth]{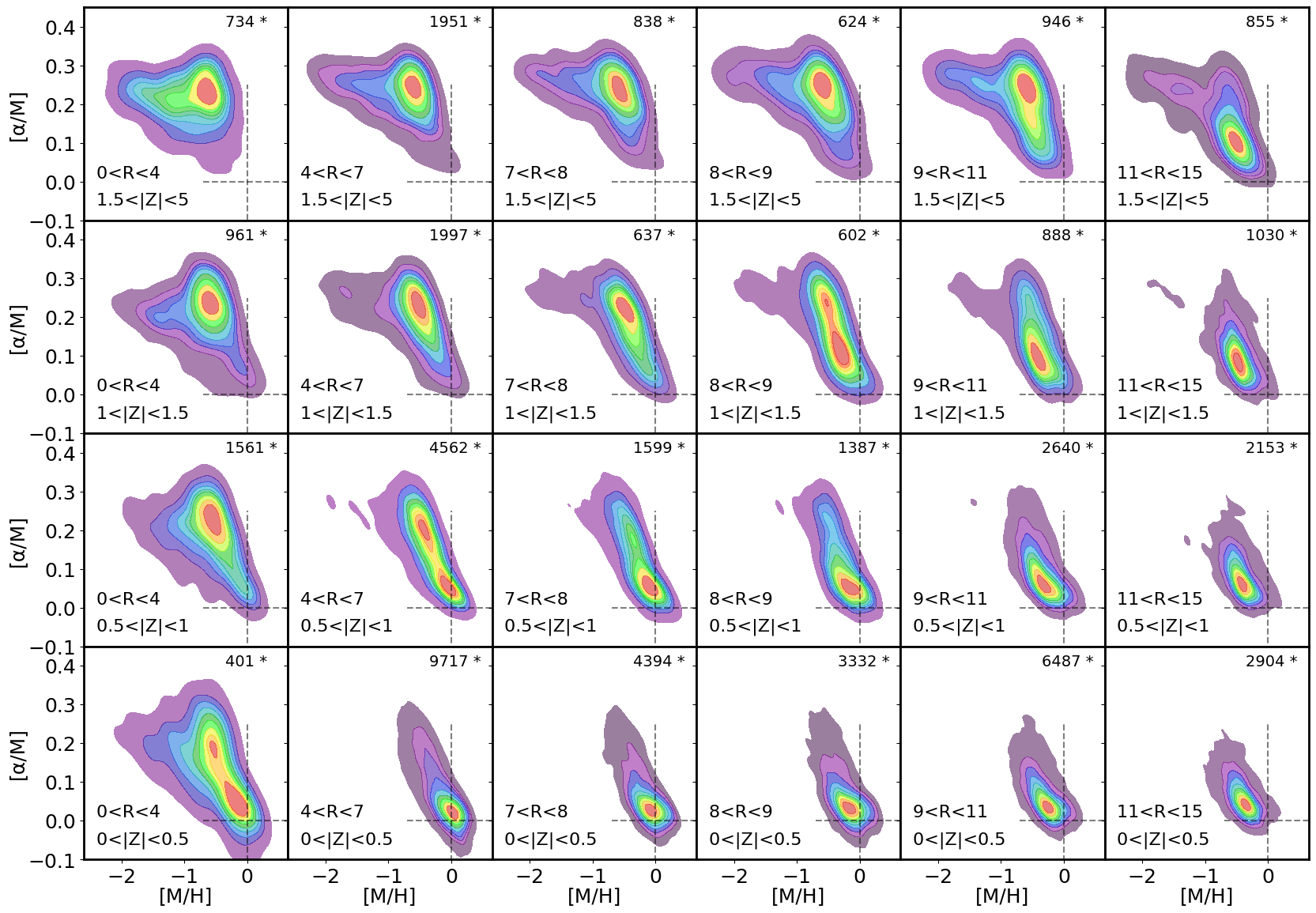}
        \caption{Two-dimensional histograms and contours of $\alpham\, vs.\,\mh$ in 53\,200 \emph{Gaia}-RVS giants
 of the observed sample with $15\le\snr\le25$ and $\logg\le2.2$ within the training sample limits. The stars are plotted in kiloparsec bins of galactocentric radius (R)
 and height above the galactic plane (Z).}
        \label{fig:queiroz_plot}
\end{figure*}

We explored the $\alpham-\mh$ bimodality for $15\le\snr\le25$,
that is, the S/N regime for which GSP-Spec does not provide $\alpham$ ratios with good quality flags.
In \figurename~\ref{fig:queiroz_plot}, we show the $\alpham-\mh$ plane decomposed
into bins of galactocentric radius R and height above the Galactic plane Z. Our sample consists of $53\,200$ stars with $15\le\snr\le25$ and $\logg<2.3$, within the training sample limits.
At low Z (|Z|<0.5\,kpc) and $\pm1$kpc around the solar radius, we mainly probed low-$\alpham$
stars. When moving towards the inner disc, we started to populate the high-$\alpham$
sequence, and a bimodality is clearly visible in the range $0<\mathrm{R}<4\,$kpc (bottom-left panel of \figurename~\ref{fig:queiroz_plot}).
Moving to higher Z ($|\mathrm{Z}|>1.0\,$kpc), the inner disc shows a clear transition
from being low-$\alpham$ populated to high-$\alpham$ populated (stars in the bins
$0<\mathrm{R}<4\,$kpc and $1.5<|\mathrm{Z}|<4\,$kpc; top-left panels of \figurename~\ref{fig:queiroz_plot}). In the outer
disc, the stars mainly show low-$\alpham$ enrichment, which is due to disc flaring,
as first suggested by \citet{minchev_2015}. Such results are
consistent with previous works \citep{Anders2014, hayden2015}. The bimodality is
also seen in the bulge region with our CNN abundances, confirming previous results
based on APOGEE DR14 and DR16 \citep{Rojas2019, queiroz2020, Queiroz2021}. By investigating the $\alpham v.s. \mh$ pattern over a large range of galactic R and Z, we
show that CNN is able to recover the main abundance trends in the Milky Way over a
large galactic volume, even for low S/N ratio RVS data.
This first detection of
the bimodality in the RVS data using a CNN is a step forward in the scientific
output of the \emph{Gaia} mission. A detailed discussion of the
candidate-bulge stars seen in this sample will be discussed in Nepal et al.
(in prep). Other interesting results will also be discussed in forthcoming papers
using the new StarHorse method run on this sample (Nepal et al., submitted).

\section{Caveats}\label{Caveats}

Stars with $\mh$ below $-2.3$ may suffer from systematics due to low statistics
    in the training sample at very low metallicities. For the future use of the CNN and
    \emph{Gaia-}RVS, a proactive training sample should be adopted by further populating
    the metal-poor regime. We are going in such a direction with an
    accepted SDSS-V open-fiber programme (PI G. Guiglion) that will observe 4\,000
    RAVE metal-poor stars \citep{matijevic2017, guiglion2020} with the
    APOGEE spectrograph. In the future, such metal-poor stars will complement the
    current training sample and improve the reliability of the metallicity
    measurements of stars below $-2.3$ dex. 
    
    Over the full RVS catalogue of $886\,080$ stars parameterised by the CNN,
    $22\%$ belong outside of the training sample limits. It is clear
    that in the present study, the performances of our CNN approach
    are limited by the training sample. The APOGEE and \emph{Gaia} surveys are
    characterised by different selection functions. The selection function of
    the training sample is then characterised by traits common to both surveys,
    but a full analysis of the selection function is beyond the scope of the present paper.
    For future \emph{Gaia} releases,
    substantial effort should be applied to populating the training sample with more
    diverse targets, such as OB stars, M dwarfs, and giants. Also, nearby stars
    (with large parallaxes) or bright stars should be added in the training
    set in order to have a more complete representation of the local stellar
    populations. Such statements are valid for ongoing and future large
    spectroscopic surveys that may want to use ML algorithms for
    spectral parameterisation, such as GALAH, SDSS-V, and 4MOST.


\section{Conclusion}\label{conclusiooooonnnn}

In June 2022, the \emph{Gaia} consortium released data of one million RVS stars, with one-third having a low signal-to-noise ratio (15$<$S/N$<$25). In this paper,
we derived atmospheric parameters and chemical abundances from this dataset
by combining, for the first time, \emph{Gaia}-RVS spectra, photometry (G, 
$\mathrm{G}\_{\mathrm{BP}}$, $\mathrm{G}\_{\mathrm{RP}}$),
parallaxes, and \emph{Gaia} XP coefficients. We summarise our method and main
achievements below:

\begin{itemize}

\item Benefitting from the last data release of the APOGEE survey, we
built a training sample with high-quality labels, including atmospheric parameters $\teff$,
$\logg$, and $\mh$ and chemical abundances $\feh$ and $\alpham$. After careful use of
\emph{Gaia} and APOGEE flags, the resulting training sample was composed of 44\,780
stars (Sect.~\ref{training_sample}, \figurename~\ref{fig:training_sample_kiel})
with RVS spectra, photometry, astrometry, and XP data. We also assembled a set
of RVS spectra (with additional photometry, astrometry, and XP data)
for which we measured the above-mentioned labels. This observed set
is composed of 841\,300 stars. 

\item We built a CNN based on previously used
architectures from \citet{guiglion2020, Nepal2023, Ambrosch2023} that we
optimised for the \emph{Gaia} datasets used in this work (Sect.~\ref{cnn_method},
\figurename~\ref{fig:cnn_architecture}). We trained a series of 28 CNN models
that we combined in order to determine average labels. We showed that the CNN learns from
relevant spectral features for a given label as well as from XP
coefficients (Sect.~\ref{section_gradients}, \figurename~\ref{fig:grad_rvs},
and \figurename~\ref{fig:gradients_XP}). We confirmed that XP coefficients
can be used for constraining atmospheric parameters as well as $\alpham$.

\item  We derived realistic uncertainties by combining the model-to-model dispersion
with the departure from the training sample input labels (Sect.~\ref{errors_section},
\figurename~\ref{fig:uncertainty_plot_training}). The uncertainties in the
observed sample are on the order of 50$-$70\,K in $\teff$; 0.1 dex in $\logg$;
0.07/0.15\,dex in $\mh$ and $\feh$; and 0.02/0.04 in $\alpham$.

\item The CNN shows a stable performance across a large range
of S/N (Sect.~\ref{stability_snr}, \figurename~\ref{fig:snr_stability}).
The dispersion with respect to APOGEE is constant with S/N
in both training
and observed samples and more robust than the GSP-Spec parameters when compared to APOGEE.
We demonstrated that the CNN is capable of precisely and accurately
parameterising metal-poor stars in the range $15<\mathrm{S/N}<25$ (see Sect.~\ref{section_metal_poor_observed}, 
and \figurename~\ref{fig:metal_poor_observed}). Such high-quality parameterisation 
is only achievable when combining spectra, photometry, parallaxes, and XP data.

\item Compared with the precise asteroseismic $\logg$ of red giant stars computed
using asteroseismic parameters from \citet{Zinn2022}, the CNN shows no mean bias or
residual trends, with a typical dispersion of 0.1 dex, which is remarkable
(Sect.~\ref{Validation_seismo}, \figurename~\ref{fig:seismic_comp}).
Such a precision can only be achieved
thanks to the external data we used in the form of photometry, parallaxes, and
XP coefficients. Comparisons with GALAH DR3 also showed the CNN to have a higher precision compared to GSP-Spec (\figurename~\ref{fig:galah_comparison}). We also
showed that CNN is very robust regarding radial velocity uncertainties
(\figurename~\ref{fig:vrad_plot}).

\item Using the dimensionality-reduction algorithm t-SNE, we classified
the RVS spectra into training-like and training-unlike spectra, allowing us to discard RVS spectra
that are not similar to the training sample. As a result, among the 841\,300
RVS stars of the observed sample, 644\,287 stars (including 10\,718 metal-poor stars)
are within the training sample limits and characterised by flag\_boundary="00000000",
and they are recommended for science applications.

\item With our dataset, it is possible, for the first time, to resolve and trace the $\alpham$-$\mh$
bimodality in the Milky Way disc using \emph{Gaia} data (Sect.~\ref{science_verification},
\figurename~\ref{fig:queiroz_plot}). Such a performance has been achieved
thanks to a large-enough and high-quality training sample combined with a
complex CNN architecture, and most important is the combining of four unique
datasets (RVS spectra, photometry, parallaxes, and XP coefficients).

\end{itemize}

As RVS spectra are rich in spectral lines, we plan to measure more
elemental abundances for the next releases of \emph{Gaia}-RVS data, such
as Ti, Si, and Ce. In addition, the next \emph{Gaia} data release will
consist of 66 months of data (expected by the end of 2025) and will include all
epoch and transit data for all sources (i.e. low S/N RVS data). This current
paper represents a step forward in the analysis of such a dataset.
For the next studies and generation of surveys, the training sample should be
built in
a proactive way, that is, by selecting targets to be observed instead of
simply using an existing set of reference stars. In this way, the biases inherent to any training sample will be limited. For instance,
focus should be on the tail of metal-poor stars as well as bright stars,
local stars, and M giants. Such a challenge will have to be faced by 4MOST, which
aims at using ML tools for stellar parameterisation.
On that topic, we believe that the experience gained here with the analysis
of spectra with limited resolution and spectral coverage will be crucial
in the development of a CNN method for future surveys, such as the 4MOST
Milky Way Disc and Bulge Low-Resolution Survey (4MIDABLE-LR; \citealt{chiappini2019}).


\begin{acknowledgements}
G.G. sincerely thanks Gal Matijevi\v c for the great time spent in Innsbruck
where at least 40\% of this paper was written. G.G. is grateful to Rene Andrae
for the fruitful discussions during the MPIA Galaxy \& Cosmology Retreat 2023.
G.G. thanks Tristan Cantat-Gaudin for providing the list of open cluster
members within the \emph{Gaia}-RVS sample. The authors acknowledge the anonymous referee for the
comments and suggestions that improved the readability of the paper. G.G. acknowledges support by
Deutsche Forschungs-gemeinschaft (DFG, German Research Foundation) – project-IDs:
eBer-22-59652 (GU 2240/1-1 "Galactic Archaeology with Convolutional Neural-Networks:
Realising the potential of Gaia and 4MOST"). G.G. and S. N. thank the SOC and LOC of the
conference "The Milky Way Revealed by Gaia: The Next Frontier" (5-7 Sept. 2023) where
results of this paper where presented. 
We thank the E-science \& IT team for COLAB service and research infrastructure at AIP.
S.N. thanks the members of the StarHorse team, Chemodynamical group and MWLV \& DGGH
sections at AIP for lively discussions. We acknowledge the International Space Science
Institute for supporting the ISSI team 542 on "CatS - A reference catalogue for
Spectroscopic surveys" led by G. Guiglion and M. Valentini, in which this paper was
partly designed. M.B. is supported through the Lise Meitner grant from the Max Planck
Society. C.C. acknowledges Fundacion Jesus Serra  for its great support during her visit
to IAC, Spain, during which part of this work was written. We acknowledge support by
the Collaborative Research Centre SFB 881 (projects A5, A10), Heidelberg University,
of the Deutsche Forschungsgemeinschaft (DFG, German Research Foundation) and by the
European Research Council (ERC) under the European Union’s Horizon 2020 research and
innovation programme (grant agreement 949173). G.T. acknowledges financial support
from the Slovenian Research Agency (research core
funding No. P1-0188). J.M. acknowledges support from the ERC Consolidator Grant funding
scheme (project ASTEROCHRONOMETRY, https://www.asterochronometry.eu, G.A. n. 772293).
This work has made use of data from the
European Space Agency (ESA) mission {\it Gaia}
(\url{https://www.cosmos.esa.int/gaia}), processed by the {\it Gaia}
Data Processing and Analysis Consortium (DPAC,
\url{https://www.cosmos.esa.int/web/gaia/dpac/consortium}). Funding for the DPAC
has been provided by national institutions, in particular the institutions
participating in the {\it Gaia} Multilateral Agreement.
This work made use of \texttt{overleaf}
\footnote{\url{https://www.overleaf.com/}}, and of the
following \textsc{python} packages (not previously mentioned):
\textsc{matplotlib} \citep{Hunter2007}, \textsc{scikit-learn} \citep{scikit-learn}, \textsc{numpy}
\citep{Harris2020}, \textsc{pandas}
\citep{mckinney-proc-scipy-2010}, \textsc{seaborn}
\citep{Waskom2021}. This work also benefited from
\textsc{topcat} \citep{Taylor2005}.
\end{acknowledgements}

\bibliographystyle{aa}
\bibliography{cite_r_s}



\appendix

\section{Details on t-SNe classification}\label{tsne_appendix}

\begin{figure*}[h]
        \centering
        \includegraphics[width=1.0\textwidth]{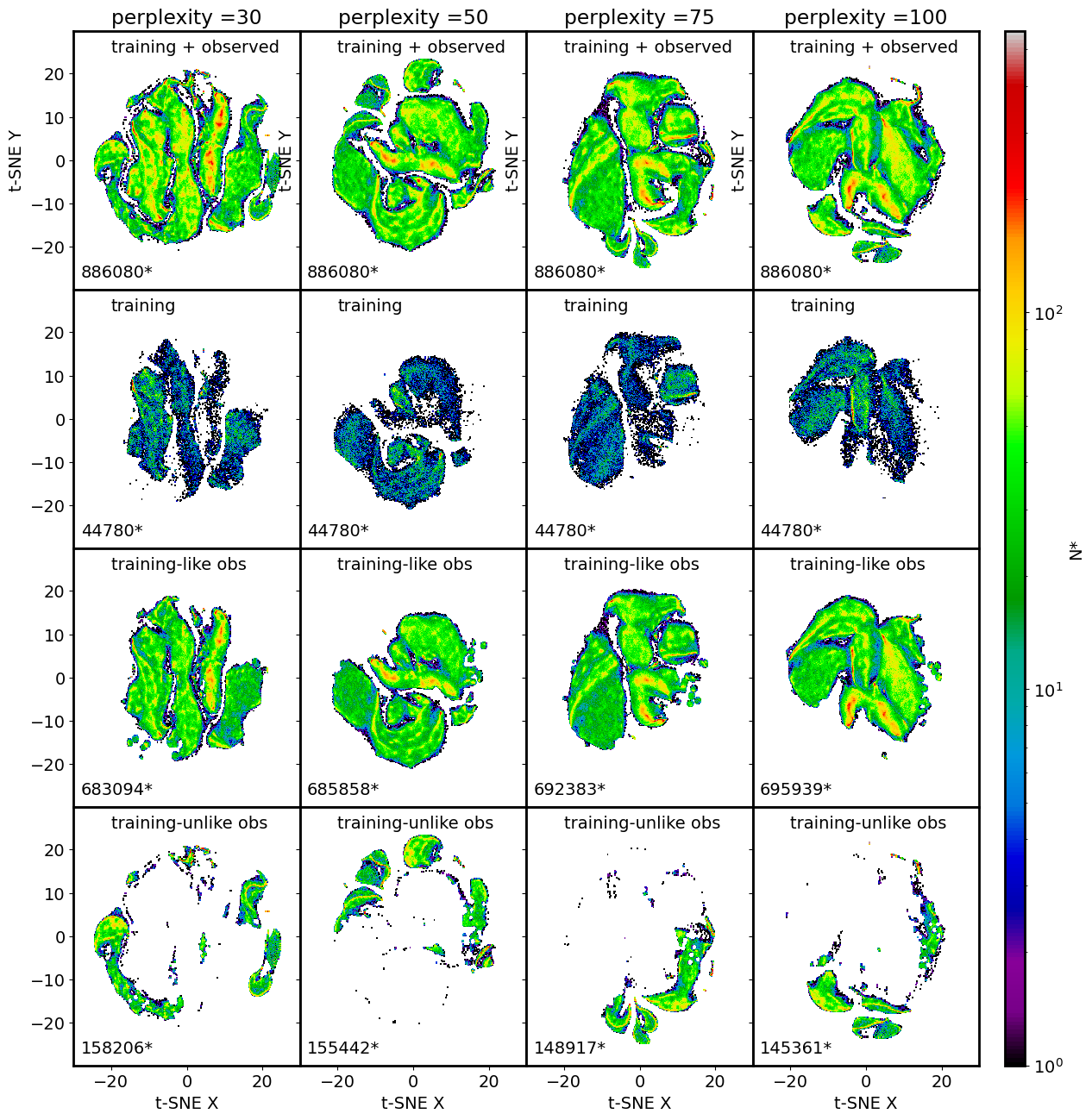}
        \caption[]{t-SNE maps of the RVS spectra for four different perplexities (4 columns: 30, 50, 75, 100). The top row shows the maps for the whole RVS sample; the second shows only the RVS classification for the training sample; and the third and fourth rows show the `training-like' and `training-unlike' spectra from the observed sample, respectively.}
        \label{fig:tsne}
\end{figure*}

We present here a comprehensive view of the t-SNE classification of RVS spectra 
for four different perplexities: 30, 50, 75, 100. Results are shown in \figurename~\ref{fig:tsne}. Overall, the t-SNE look visually similar, with quite consistent numbers of `training-like' and `training-unlike' spectra from the observed sample.

\begin{figure*}[h]
        \centering
        \includegraphics[width=1.0\textwidth]{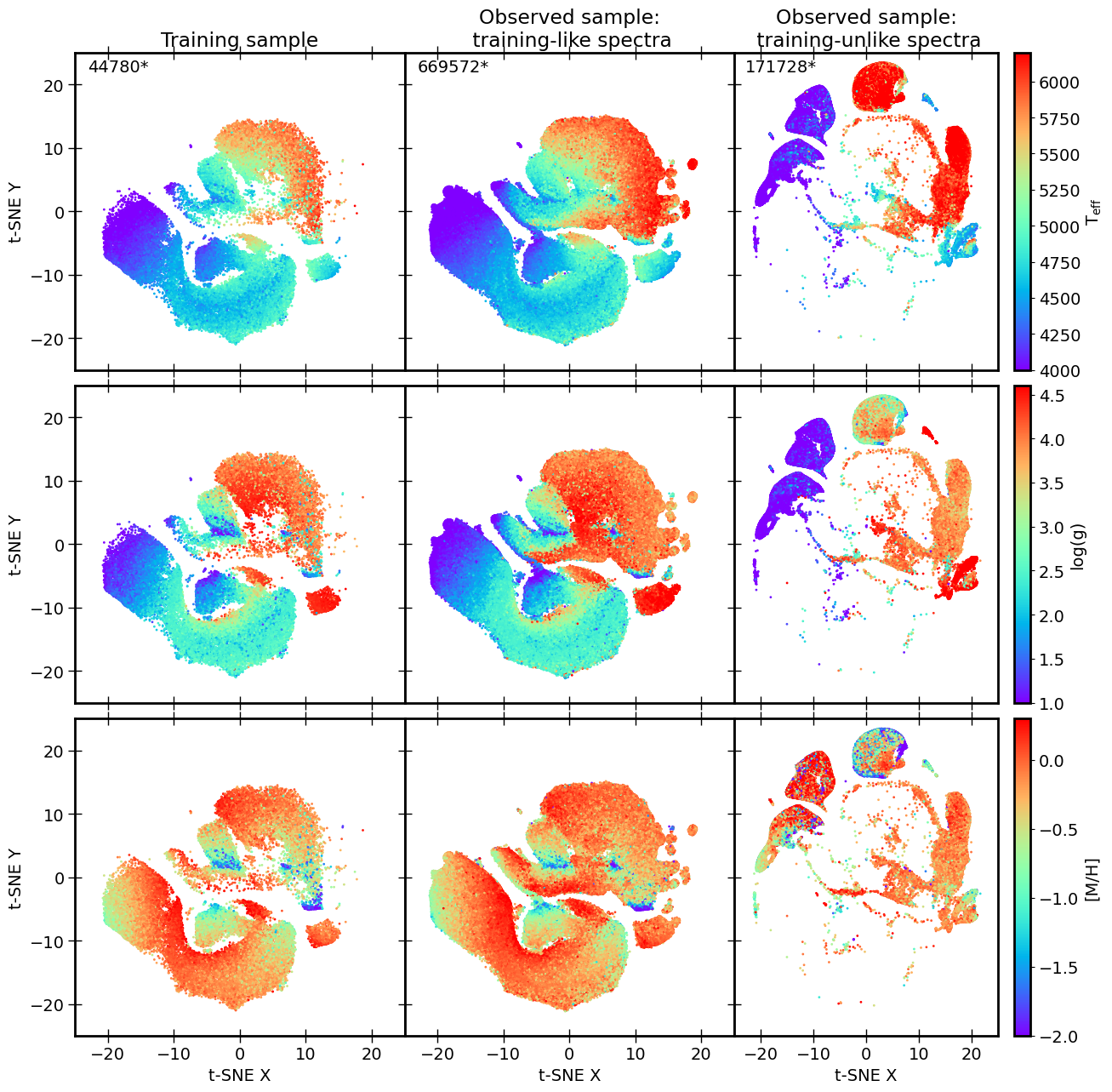}
        \caption{t-SNE maps for perplexity=50 for the training sample spectra (left) and the `training-like' (middle) and `training-unlike' (right) spectra of the observed sample. The maps are colour-coded with $\teff$ (top), $\logg$ (centre), and $\mh$ (bottom).}
        \label{fig:tsne_teff_logg_mh}
\end{figure*}

To give the reader an idea of how the t-SNE classification correlates with atmospheric parameters, we present in \figurename~\ref{fig:tsne_teff_logg_mh} the t-SNE classification with perplexity=50 colour-coded with $\teff$, $\logg$, and $\mh$. One can clearly see in the `training-unlike' map blobs of cool and hot stars that are not present in the `training-like' map. Such blobs correspond to the cool giants and hot dwarfs described in Section~\ref{explore_labels_lim} and were clearly mislabelled by the CNN. Using this t-SNE approach allows us to adequately flag the mislabelled spectra and then provide the user a robust flag for cleaning the CNN catalogue of spurious measurements.

\section{Training convolutional neural network of purely \emph{Gaia}-RVS spectra}\label{only_rvs}

\begin{figure}[h]
        \centering
        \includegraphics[width=1\linewidth]{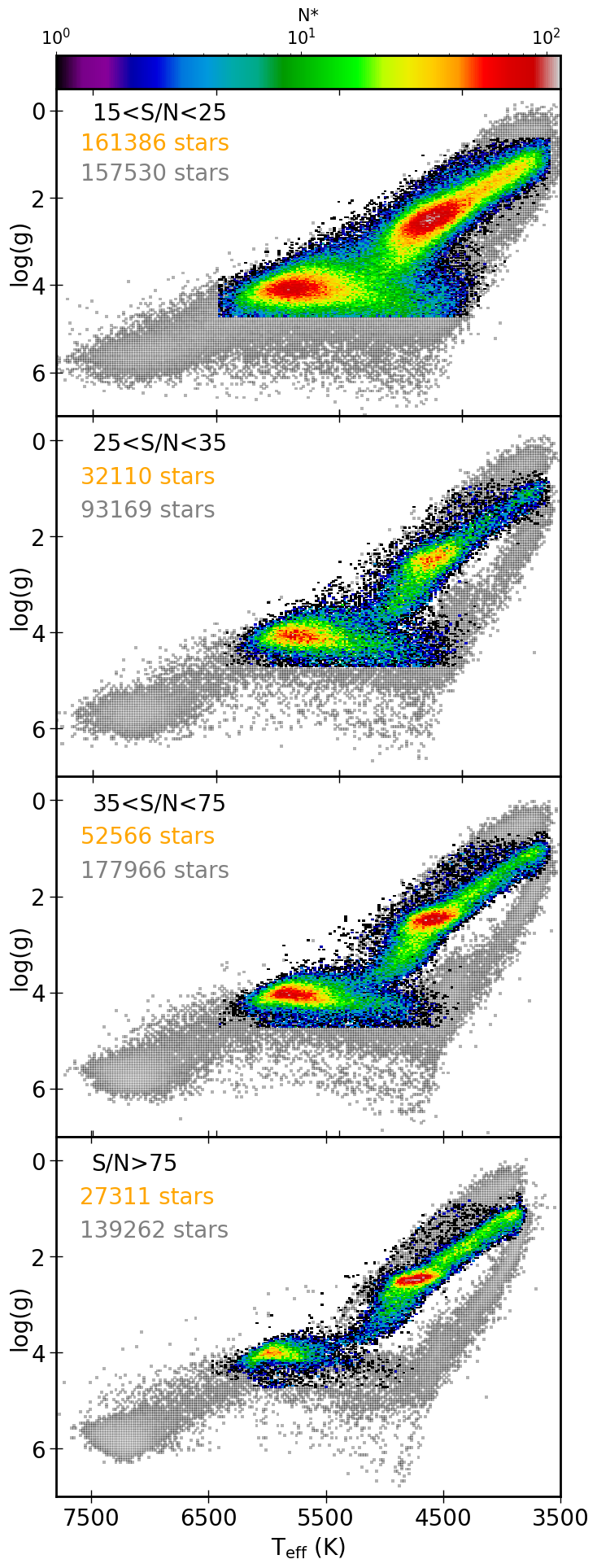}
        \caption[]{Same figure as \figurename~\ref{fig:obs_snr_cut} but for RVS stars 
 parameterised by CNN only using RVS spectra.}
        \label{fig:obs_snr_cut_no_phot}
\end{figure}

We have demonstrated in the previous sections that combining
RVS spectra, magnitudes, parallaxes, and XP coefficients
allowed us to provide high-quality CNN labels. For completeness, we trained 
CNN using only RVS spectra as input data. In
\figurename~\ref{fig:obs_snr_cut_no_phot}, we show Kiel diagrams of the
high-quality sample in bins of S/N (15-25, 25-35, 35-75, and $\ge75$),
colour-coded with $\mh$ (219\,145 stars within training set limits.
We present the rest of the RVS sample in grey (stars outside
the training sample limits or with large uncertainties). We observed a
consistent Kiel diagram, even at low S/N, with a clear metallicity
sequence in the giant branch, while the red clump locus is clearly
reproduced. The sequence of cool dwarfs only extends down to $\teff=5\,000\,$K,
contrary to \figurename~\ref{fig:obs_snr_cut}. This is due to the fact that,
similar to RAVE spectra, RVS seems to suffer from degeneracies as well. 
Such degeneracies are clearly brought to light by the stars in grey, that is, a sequence 
connecting the cool giants to the cool dwarfs (see
\citealt{kordopatis2013, guiglion2020} for more details of the RAVE spectral 
degeneracies). Another interesting feature is 
a high concentration of stars above $\teff>6500\,$K. Their spectra are characterised 
by strong Hydrogen Paschen lines. The CNN seems to constrain their 
temperature rather well, while $\logg$ is clearly biased.

\begin{figure}[h]
        \centering
        \includegraphics[width=1\linewidth]{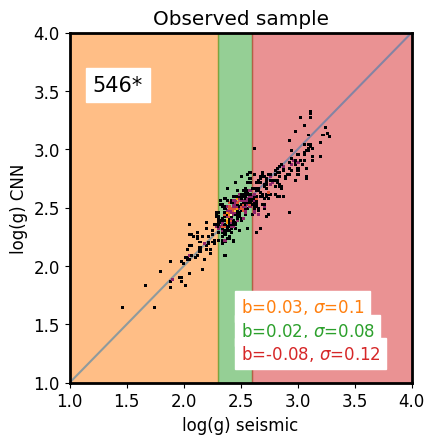}
        \caption{Same figure as \figurename~\ref{fig:seismic_comp} but comparing
 the CNN gravities (derived using only RVS spectra) and seismic gravities}.
        \label{fig:seismic_comp_no_phot}
\end{figure}

In \figurename~\ref{fig:seismic_comp_no_phot}, we compare CNN gravities (derived 
from RVS spectra only) to seismic gravities computed from \citet{Zinn2022}
(see Sect.~\ref{Validation_seismo}; same stars as in \figurename~\ref{fig:seismic_comp}).
Overall, the bias is well below 0.1 dex, while the dispersion ranges
from 0.08 to 0.12\,dex, which is remarkable considering that we only
trained on RVS spectra. We emphasise that calibrated GSP-Spec gravities only showed 
a precision of 0.14-0.17\,dex compared to seismic gravities.

Through such tests, we show that the CNN still provides reliable parameters
when trained only on RVS spectra. This is due to the high-quality APOGEE labels
transferred and learned during the training process. The CNN is still not immune
to spectral degeneracies for some stars. Such degeneracies are broken when using extra magnitudes,
parallaxes, and XP data.


\section{Convolutional neural network stability with respect to \emph{Gaia} DR3 radial velocities}\label{vrad_test}

\begin{figure*}[h]
        \centering
        \includegraphics[width=1.0\linewidth]{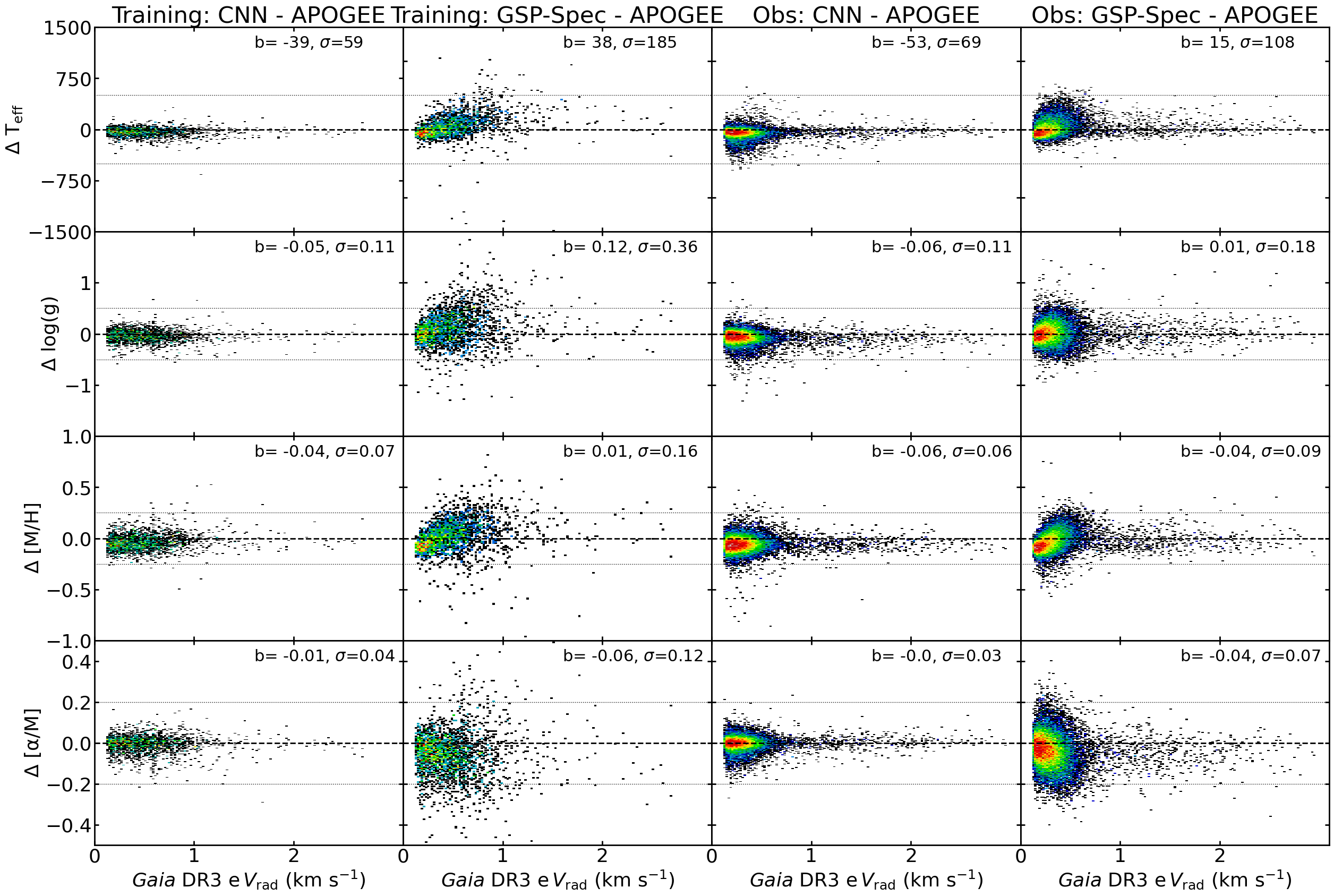}
        \caption{Two-dimensional density distribution of the CNN minus APOGEE and
 calibrated GSP-Spec minus APOGEE as a function of \emph{Gaia} DR3
 radial velocity uncertainties in both the training (2\,606 stars)
 and observed samples (20\,948 stars, within the training sample limits).}
        \label{fig:vrad_plot}
\end{figure*}

In this section, we investigate how sensitive the CNN is to the radial velocity uncertainties. In this study, the adopted
\emph{Gaia}-RVS have been corrected from Doppler shift by the
\emph{Gaia} consortium. It may happen that a tiny residual shift
(a fraction of a pixel) can be present in the corrected spectra of stars 
with large Vrad uncertainties. We note that small systematics in the radial
velocity applied to spectra may lead to systematics in the predicted
labels (see \citealt{Nepal2023}).
In \figurename~\ref{fig:vrad_plot}, we show how the difference
in labels between CNN and APOGEE and GSP-Spec and APOGEE (in both training and
observed samples) varies with \emph{Gaia} DR3 radial velocity uncertainties
(e$\mathrm{V}_{\mathrm{rad}}$). We applied the recommended flags\_gspspec
in order to clean the GSP-Spec sample. First, for the training sample, we clearly observed
that for each label ($\teff$, $\logg$, $\mh$, and $\alpham$), the systematics
(bias) between CNN and APOGEE is constant as a function of
e$\mathrm{V}_{\mathrm{rad}}$. On the other hand, GSP-Spec\,-\,APOGEE shows
strong residual trends for the bulk of the distribution, with increasing
bias as a function of e$\mathrm{V}_{\mathrm{rad}}$, as documented in \citet{recioblanco2023}.
The dispersion is also
two to three times larger compared to the CNN. For the stars on the observed sample, we
observed similar systematic trends. We note that even if not plotted here,
$\feh$ behaves similarly as $\mh$. With such tests, we can conclude that
the CNN shows no strong sensitivity to e$\mathrm{V}_{\mathrm{rad}}$ and shows
no residual trends. 

\end{document}